\documentclass[ 
preprint,aps,prx,superscriptaddress,twocolumn,longbibliography,notitlepage,nofootinbib, 10pt]{revtex4-2}

\usepackage[breaklinks=true,colorlinks,citecolor=color1,linkcolor=color1,urlcolor=color1]{hyperref}
\usepackage{graphicx}
\usepackage{dcolumn}
\usepackage{bm}
\usepackage{color}
\definecolor{color1}{rgb}{0,0,1}
\definecolor{color3}{rgb}{0.85,0,0.5}
\usepackage{graphicx}
\usepackage{amsmath}
\usepackage{amssymb}
\usepackage{amsfonts}
\usepackage{amsthm}
\usepackage{bbm}
\usepackage{soul}
\usepackage{textcomp}
\usepackage{float}
\usepackage{balance}
\usepackage{bm}
\usepackage{dsfont}
\usepackage{relsize}
\usepackage{braket}
\usepackage{enumitem}
\usepackage{cleveref}
\usepackage{float}
\usepackage{siunitx}
\usepackage{xcolor}
\usepackage{booktabs}
\usepackage[makeroom]{cancel}

\usepackage{textgreek}

\DeclareSIUnit\bar{bar}
\newcommand{\sfref}[2]{\textcolor{color1}{\hyperref[#1]{Fig.$\,$\ref{#1}(#2)}}}
\newcommand{\fref}[1]{\textcolor{color1}{\hyperref[#1]{Fig.$\,$\ref{#1}}}}

\newcommand{\aref}[1]{\textcolor{color1}{\hyperref[#1]{App.$\,$\ref{#1}}}}
\newcommand{\tref}[1]{\textcolor{color1}{\hyperref[#1]{Table$\,$\ref{#1}}}}

\definecolor{color1}{rgb}{0,0,0.7}
\definecolor{color2}{rgb}{0.85,0,0}
\begin{document}


\title{\textbf{Fabrication-free assessment of microwave losses in germanium-based dielectrics and superconductors} 
}%

\author{Haoran Lu}
\thanks{These authors contributed equally}
\email{hl2396@cornell.edu}
\affiliation{School of Applied and Engineering Physics, Cornell University, Ithaca, NY, 14853, USA}

\author{Kushagra Aggarwal}
\thanks{These authors contributed equally}
\email{ka543@cornell.edu}
\affiliation{School of Applied and Engineering Physics, Cornell University, Ithaca, NY, 14853, USA}

\author{Xiangqin Wang}
\affiliation{School of Applied and Engineering Physics, Cornell University, Ithaca, NY, 14853, USA}

\author{Pauline Drexler}
\affiliation{Fakultät für Physik, Universität Regensburg, 93040 Regensburg, Germany}

\author{Daniel Tong}
\affiliation{School of Applied and Engineering Physics, Cornell University, Ithaca, NY, 14853, USA}
\affiliation{Department of Materials Science and Engineering, Cornell University, Ithaca, NY, 14853, USA}

\author{Maciej W. Olszewski}
\affiliation{Department of Physics, Cornell University, Ithaca, NY 14853, USA}%

\author{Anand Ithepalli}
\affiliation{Department of Materials Science and Engineering, Cornell University, Ithaca, NY, 14853, USA}

\author{Lingda Kong}
\affiliation{School of Applied and Engineering Physics, Cornell University, Ithaca, NY, 14853, USA}

\author{Debdeep Jena}
\affiliation{Department of Materials Science and Engineering, Cornell University, Ithaca, NY, 14853, USA}
\affiliation{Kavli Institute at Cornell for Nanoscale Science, Cornell University, Ithaca, NY 14853, USA}
\affiliation{School of Electrical and Computer Engineering, Cornell University, Ithaca, NY 14853, USA}

\author{Peter L. McMahon}
\affiliation{School of Applied and Engineering Physics, Cornell University, Ithaca, NY, 14853, USA}
\affiliation{Kavli Institute at Cornell for Nanoscale Science, Cornell University, Ithaca, NY 14853, USA}

\author{David A. Muller}
\affiliation{School of Applied and Engineering Physics, Cornell University, Ithaca, NY, 14853, USA}
\affiliation{Kavli Institute at Cornell for Nanoscale Science, Cornell University, Ithaca, NY 14853, USA}

\author{Dominique Bougeard}
\affiliation{Fakultät für Physik, Universität Regensburg, 93040 Regensburg, Germany}

\author{Valla Fatemi}
\email{vf82@cornell.edu}
\affiliation{School of Applied and Engineering Physics, Cornell University, Ithaca, NY, 14853, USA}

\date{\today}%
\begin{abstract}

We present a flip-chip-based sensing scheme to measure effective microwave losses associated with target materials for quantum technologies, without requiring any device fabrication on the material under test. 
Using this approach, we quantify the microwave losses of a strain-engineered Ge/SiGe quantum well heterostructure and investigate losses arising from its Ge substrate and intermediate layers.
The quality factors of the fabricated microwave resonators agree with the losses of dielectric materials independently extracted from flip-chip sensing measurements.
We further study the superconductor platinum silicon germanide (PtSiGe) prepared by thermal reaction with a deposited Pt film, finding high microwave losses that limit the suitability of the films studied here as the sole superconductor for high-quality resonator applications. 
By coating Pt with Nb prior to the reaction, we observe a substantial reduction in microwave loss and a nearly three-fold enhancement of the transport critical temperature.
The temperature dependence of the microwave loss is consistent with gap inhomogeneity in both superconducting films.
These results identify constraints on material choices, provide design guidance for microwave circuits on planar Ge heterostructures, and demonstrate a fast-turnaround testing method for new materials for superconducting quantum circuits.

\end{abstract}

\maketitle


\section{Introduction}
Germanium-based heterostructures have emerged as a leading platform for spin qubits and hybrid superconductor-semiconductor devices~\cite{Scappucci2020}. 
Ge two-dimensional hole gases (2DHGs) in particular offer a combination of favorable properties such as low effective mass, weak hyperfine interaction, strong spin-orbit interaction, and high mobility. 
Such planar Ge quantum wells have demonstrated significant progress toward scalable, electrically controllable spin qubits~\cite{Hendrickx2020, Hendrickx2021, Jirovec2021}. 
Moreover, highly transparent superconducting contacts to planar Ge have been achieved, making this platform an attractive host for superconductor-semiconductor hybrid devices~\cite{sagi_gate_2024,kiyooka_gatemon_2025, pitav-vidal_review_2025, lakic_quantum_2025, Ruggiero_prerprint_2026}, such as Andreev spin qubits~\cite{hays_coherent_2021, pita-vidal_direct_2023}.

Such Andreev-based and other hybrid architectures require superconductors that simultaneously form highly transparent contacts to the 2DHG and low-loss microwave circuitry for qubit readout, control, and long-range coupling. 
While substantial progress has been made toward achieving high-transparency contacts in planar Ge~\cite{valentini_parity_2024, Vigneau2019, Tosato2023, Aggarwal2021}, the realization of high-quality microwave resonators on this platform has proven more challenging~\cite{valentini_parity_2024, Nigro_loss_QW_2024, Ruggiero2026}.
This is because fully fabricated microwave resonators on planar Ge typically exhibit high losses~\cite{Nigro_loss_QW_2024, valentini_parity_2024, Ruggiero2026, Palma_preprint_2025}.

The origins of these losses remain poorly understood. 
Microwave losses may come from a combination of the substrate, the deposited dielectric layers, the choice of superconducting material, the interfaces between these layers and their vacuum-facing surfaces~\cite{mcrae_materials_2020,de_leon_materials_2021}. 
Furthermore, the intrinsic losses of the materials and surfaces can be degraded due to processing-induced damage or contamination during fabrication~\cite{olszewski_low_2025,Gao2022,sandberg_etch_2012}.
Disentangling these different contributions is essential for the systematic development of scalable microwave architectures on planar Ge. 
Existing methods to disentangle these different sources rely on fabricating a series of different microwave resonator geometries from the materials of interest to tailor how the resonator electric field samples different material regions~\cite{sage_study_2011,woods_determining_2019}, but this is a slow and fabrication-development-intensive approach. 
Fabrication-free approaches have been explored for large-volume substrates~\cite{Read_precision_2023} but not yet for thin film dielectrics and superconductors.
A fast-turnaround method with a minimum number of fabrication steps, both for dielectric materials and superconducting thin films, will be beneficial for speeding up the research cycle and minimizing confounding factors, particularly for newer materials with less prior investigation. 

Here, we introduce a flip-chip sensing technique that constrains microwave losses associated with materials without requiring device fabrication on the material under test.
We apply this method to Ge-based heterostructures and superconductors to constrain sources of microwave loss in dielectrics and superconducting thin films.
We find that the Ge wafer and the strain-relaxing, composition-graded SiGe buffer layer exhibit similar properties.
Further, our results reveal that PtSiGe exhibits substantial microwave losses that limit its utility as a standalone superconducting film for resonator applications; we show that these losses may be mitigated by heterostructuring with low-loss superconductors.
Our results provide practical guidance on material choice and device design strategies, and offer a route to disentangling the microscopic sources of microwave loss needed for planar Ge quantum technology and other solid-state quantum technologies.
Our approach helps constrain dominant dissipation mechanisms and provides a general platform for rapidly screening materials for quantum technologies.

\begin{figure}[htb]
\centering
\includegraphics[width = \columnwidth]{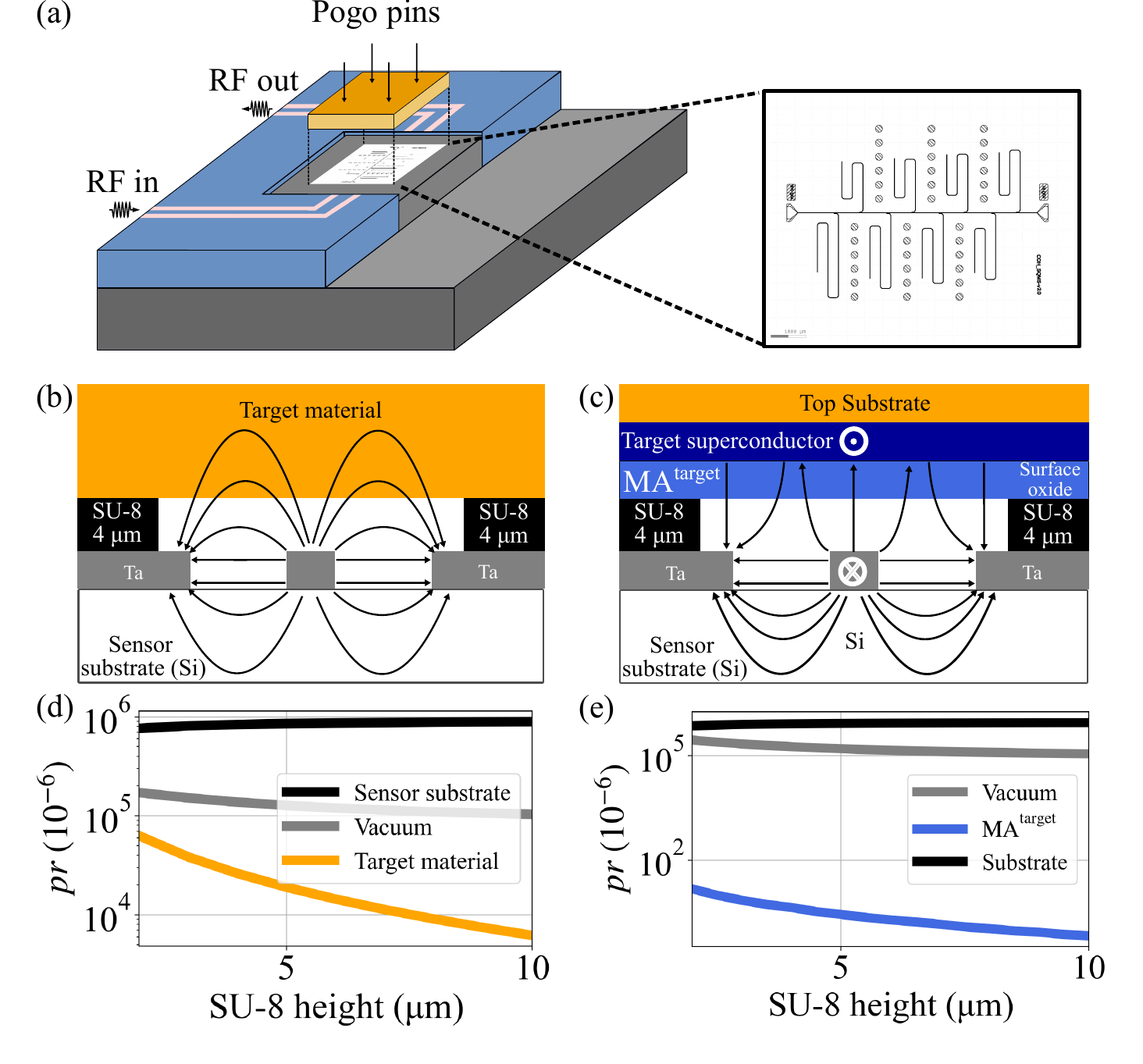}
\caption{ 
\textbf{Sensing scheme.} 
(a) Schematic of the flip-chip-based sensing scheme.
The sensor chip, with microwave resonators and \SI{4}{\micro\meter} SU-8 pillars, is bonded to the package using Al wire.
Target chips are placed above the sensor chip, and tight contact is ensured by four pogo pins that press the target material from the top.
(b) Dielectric material loss sensing scheme and EM field distribution.
(c) Superconducting thin film loss sensing scheme with EM field and current distribution.
(d) Participation ratio of the dielectric target material $pr_\mathrm{material}$ as a function of the height of the SU-8 pillar, with $\epsilon_r=16.2$ of the target dielectric material.
(e) Participation ratio of target superconducting thin film surface oxide ($\text{MA}_\text{target}$) as a function of the height of the SU-8 pillar, for a target film with negligible kinetic inductance. 
The target film surface oxide is assigned $\epsilon_r=10$ and of \SI{1}{\nano\meter} thickness.
} 
\label{fig:scheme} 
\end{figure}

\section{Flip-chip technique for sensing microwave properties of materials}

We developed a flip-chip-based superconducting resonator sensor to probe the loss of different target films.
The sensing resonator is made from a \SI{100}{\nano\meter} Ta film deposited on a \SI{5}{\nano\meter} Nb seed layer on a Si substrate~\cite{olszewski_low_2025}. 
The sensing chip contains eight $\lambda/4$ CPW resonators (center metal width \SI{6}{\micro\meter}, gap width \SI{12}{\micro\meter}), with bare resonance frequencies distributed across \SIrange{4.2}{7.2}{\giga\hertz}. 
Arrays of SU-8 photoresist pillars, which are  \SI{4}{\micro\meter} tall and \SI{100}{\micro\meter} in radius, are used as standoffs for the target chip.
Further details of the resonator fabrication and SU-8 pillar patterning are included in~\aref{app:fab}.
The measurements are conducted with a mixing chamber temperature of \SI{10}{\milli\kelvin}, unless otherwise specified.

The target chip is placed on top of the device, and the package is then assembled.
Pogo pins in the package are used to maintain a compressive force between the target chip and the sensing chip, as shown in \sfref{fig:scheme}{a}.
This ensures good contact between the sample and the pillars and consequently maintains a consistent spacing between the chips.
SU-8 pillars have been reported to be robust against thermal cycling and do not deform significantly under mechanical pressure~\cite{norri_flips_2024,sainiemi_mask_2007,li_uv_2005}, and we also measured a change of less than 3\% in SU-8 pillar height after 10 thermal cycles.
The resonator frequency is sensitive to the distance between the target chip and the sensor chip, allowing verification of its presence and spacing by comparison with simulations.

The electromagnetic field of the sensing resonator extends into the target material or up to the target superconducting thin film (see \sfref{fig:scheme}{b,c}).
For dielectrics, this electric field participation in the target allows the sensor resonator to probe losses originating from both the surface and bulk of the material. 
The participation ratio of the target dielectric material $pr_\mathrm{material}$ is defined as:
\begin{equation}
    pr_{\mathrm{material}} \equiv \frac{U^\text{E}_{\mathrm{material}}}{U^\text{E}_{\mathrm{tot}}} \equiv \frac{\frac{1}{2} \int_{V_{\mathrm{material}}} \mathbf{D} \cdot \mathbf{E} d V}{\frac{1}{2} \int_{V_{\mathrm{tot}}} \mathbf{D} \cdot \mathbf{E} d V},
\end{equation}
where $U^\text{E}_{\mathrm{material}}$ is the electric energy stored in the target dielectric material and $U^\text{E}_{\mathrm{tot}}$ is the total electric energy~\cite{Read_precision_2023,wenner_surface_2011}.  
The distance between the sensing chip and target chip, controlled by the height of the SU-8 pillars, changes the $pr_{\mathrm{material}}$.
Below, we describe the sensitivity of our method to dielectrics and to superconducting films, which scales inversely with SU-8 height, depicted in \sfref{fig:scheme}{d,e}.

For Ge with a relative dielectric constant of around 16.2~\cite{emminger_temperature_2020,kopas_low_2021}, $pr_{\mathrm{material}}$ can be tuned from 6.29\% to 0.62\% by choosing the SU-8 pillar height from \SI{2}{\micro\meter} to \SI{10}{\micro\meter}, as shown in \sfref{fig:scheme}{d}.
For the \SI{4}{\micro\meter}-tall SU-8 pillars used in this work, we obtain $pr_{\mathrm{material}}=2.62\%$, which corresponds to a detectable loss tangent range of $20\times10^{-6}$ to $1000\times10^{-6}$.
We note that about 26\% of the electric energy in the target dielectric material is stored within the top \SI{2}{\micro\meter}, enabling the sensor to probe loss contributed from the heterostructure while remaining sensitive to potential microwave loss channels buried in the bulk material.
Details of the numerical simulations and dependence of the stored electric field energy on depth are described in \aref{app:pr_mat}.

For superconducting thin films, the main added contributions to the resonator loss are likely to be conductor loss and metal-air interface dielectric loss of the flipped film.
The electric energy participation ratio of the target film's metal-air (MA) surface oxide loss is $pr_\text{MA}=4.3\times 10^{-6}$, assuming an oxide layer with a thickness of \SI{1}{\nano\meter} and a dielectric constant $\epsilon_r=10$, comparable to the MA loss participation ratio of the sensing resonator $6.8\times 10^{-6}$. 
The details of the numerical simulation for estimating this can be found in ~\aref{app:MA}. 
By varying SU-8 pillar height from \SI{2}{\micro\meter} to \SI{10}{\micro\meter}, $pr_\text{MA}$ can be tuned from $14.8\times10^{-6}$ to $0.66\times 10^{-6}$, as shown in \sfref{fig:scheme}{e}.

Our sensing scheme does not require any fabrication on the target chip beyond dicing or cleaving of a chip from the source wafer. 
This contrasts with approaches that require fabrication directly on target materials, which expose surfaces to resist, etch processes, solvent baths, and general handling.
These approaches complicate the extraction of various losses~\cite{woods_determining_2019,ganjam_surpassing_2024}, especially considering that the different metal-air interfaces are not equivalent while each also contributes significantly~\cite{crowley_disentangling_2023, sandberg_etch_2012,nersisyan_manufacturing_2019}. 
Additionally, the precise details and chemical exposure history of the nanofabrication process can be important in determining microwave losses of the ultimate device~\cite{olszewski_low_2025}.
Moreover, by not requiring fabrication on novel materials, microwave loss testing requires fewer steps between heterostructure synthesis and testing.
Therefore, fabrication-minimal approaches will help to minimize confounding factors and speed up the research cycle in addressing materials challenges to microwave devices in novel materials platforms.

\section{Results}
\subsection{Dielectric material loss}

\begin{figure*}[htb]
\centering
\includegraphics[width = \textwidth]{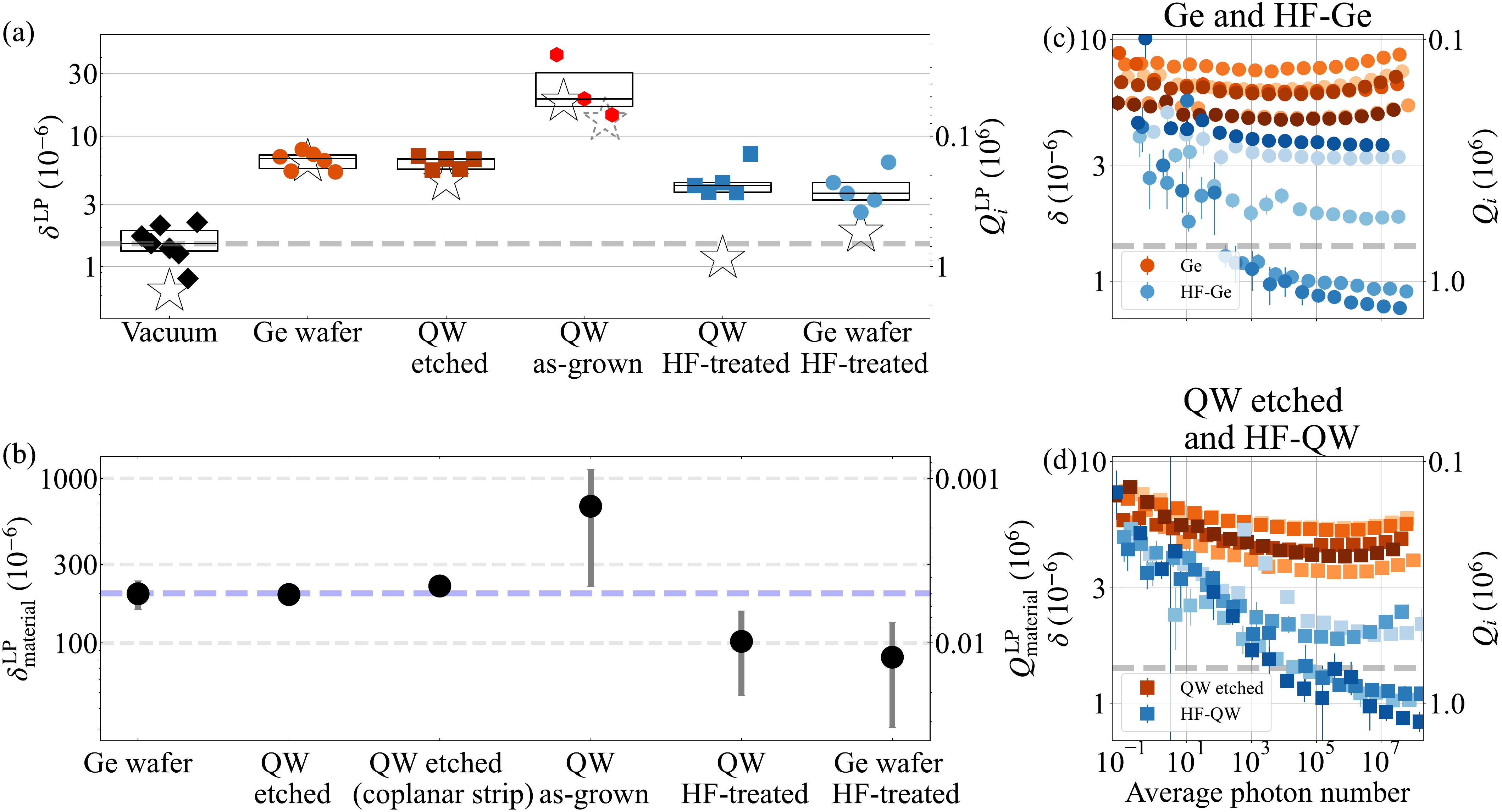}
\caption{ 
\textbf{Dielectric material loss.} 
(a) Box plots of average low-power (all $\braket{n}<10$) intrinsic loss $\delta^\text{LP}$ (dots) and average high-power (defined here as the average of the largest two photon number points from all resonators with $\braket{n} < 10^7$) intrinsic loss $\delta^\text{HP}$ (stars) for dielectric materials measured by the flip-chip-based sensor.
The dashed star of QW as-grown indicates its minimal loss, reached around $\braket{n}\approx10^4$.
The dashed grey line indicates the median loss for the ``vacuum" case, which serves as the reference loss.
(b) Inferred low power material loss $\delta^\text{LP}_\text{material}$ extracted from the measured sensor resonator loss.
The dashed blue line is set at $200\times 10^{-6}$ for visual reference.
(c) Measured loss of sensing resonator $\delta$ as a function of photon number with Ge (orange) and HF-treated Ge (blue) as the target material.
(d) Measured loss of sensing resonator $\delta$ as a function of photon number with QW-etched heterostructure (orange) and HF-treated QW heterostructure (blue) as the target material. 
}
\label{fig:material} 
\end{figure*}

We first applied the sensing scheme to probe microwave losses in three different dielectric materials:
(1) commercial epi-ready Ge substrates (room-temperature resistivity $>$\SI{30}{\ohm\centi\metre}); (2) undoped Ge/$\mathrm{Ge_{0.75}Si_{0.25}}$ rectangular quantum well (QW) heterostructures which include a strain-relaxed virtual $\mathrm{Ge_{0.75}Si_{0.25}}$ layer on the commercial Ge substrate (\aref{app:heterostructure}); (3) strain-relaxed $\mathrm{Ge_{0.75}Si_{0.25}}$ virtual layers on the Ge substrates alone, by selectively etching away the QW of the heterostructure (\aref{app:virtualSiGe}).

\sfref{fig:material}{a} presents the measured loss of the sensing resonator under these different dielectric substrates. 
The average high-power (defined here as the average of the largest two photon number points from all resonators with $\braket{n} < 10^7$) loss $\delta^\text{HP}$  is indicated by star markers, while the average low-power (all $\braket{n}<10$) loss $\delta^\text{LP}$ of each resonator is represented by colored dots.
As a reference, we also measured the intrinsic loss of the sensing resonator, which has undergone the SU-8 pillar fabrication, in the absence of any dielectric substrate (labeled “vacuum”), yielding a baseline loss of  $1.5\times 10^{-6}$.

We convert the measured loss from the flip-chip-based sensor into a sample-specific loss by subtracting the intrinsic loss of the bare sensing resonator.
Then, we normalize by the dielectric material participation ratio, $pr_\text{material}$.
This quantity is shown in \sfref{fig:material}{b}.
We note that this metric may naively be interpreted as a bulk dielectric loss, but the data contain both bulk-like and surface-like losses, both of which will be relevant (see below).
Thus, this metric may serve as an upper bound on the loss tangent of the top few microns of the sample.

For the Ge substrate and the heterostructure etched down to the $\mathrm{Ge_{0.75}Si_{0.25}}$ virtual layer (labeled hereafter as ``QW etched''), we extract low-photon loss tangents $\delta^\text{LP}_\text{material}$ of $(199 \pm 40)\times 10^{-6}$ and $(197 \pm 29)\times 10^{-6}$, respectively.\footnote{The statistical uncertainty is defined as $\sqrt{(\sigma^2_\text{target})+(\sigma^2_\text{baseline})}/pr_\text{material}$, where $\sigma$ is the standard deviation of the target material or baseline (vacuum).}
The microwave losses observed in Ge are substantially higher than those reported for bulk float-zone Si, for which loss tangents as low as $0.01\times 10^{-6}$ have been measured~\cite{acharya_ultra_2026}.

Our measurements suggest that, for the heterostructure used in this work, the strain-relaxed, composition-graded $\mathrm{Ge_{0.75}Si_{0.25}}$ virtual layer and interfaces within the heterostructure do not exhibit a substantially different loss tangent than bulk Ge at single-photon power at the present sensitivity.
To verify these results from the flip-chip sensing scheme, we fabricated a Nb coplanar strip resonator with a \SI{15}{\micro\meter} gap and a \SI{10}{\micro\meter} center trace on the same heterostructure, following removal of the QW.
At low photon number ($\langle n\rangle < 10$), we measure a total loss of $(208 \pm 19)\times 10^{-6}$. Electromagnetic simulations indicate a substrate participation ratio of $93.9\%$ for this geometry (see \aref{app:cpw}).
After accounting for participation, we extract a material loss tangent $\delta^\text{LP}_\text{material}$ of $(222 \pm 20)\times 10^{-6}$ for the SiGe substrate seen by the resonator.
This value is $10\%$ higher than the flip-chip result, demonstrating good agreement between the two measurement approaches.
The additional loss seen by the on-chip resonator measurements further suggests that, in our device, fabrication effects and metal-substrate interfaces could cause additional microwave losses relative to the dielectric losses found via the flip-chip method.

We now focus on the as-grown Ge/SiGe QW heterostructure, first without any surface treatment (labeled as ``QW as-grown'').
We found an apparent material loss tangent $\delta^\text{LP}_\text{material} = (678\pm 457)\times 10^{-6}$, such that the mean loss is about three times higher than that of pure Ge and the QW-etched heterostructure.
However, this as-grown heterostructure has a finite residual conductance at cryogenic temperatures.
An HF treatment of the as-grown heterostructure surface suppresses this residual conduction (\aref{app:fab} and \aref{app:dcdata}).
Applying this to our sample (labeled as ``QW HF-treated''), we indeed found that the loss tangent is reduced to $\delta^\text{LP}_\text{material} = (102\pm 54)\times 10^{-6}$, which is similar to the loss tangent for an HF-treated Ge substrate of $(82 \pm 51)\times 10^{-6}$.
These results indicate that the HF-treated Ge/SiGe QW heterostructure exhibits microwave performance comparable to that of single-crystalline Ge.
The treatment suppresses residual conduction in the heterostructure while partially removing surface oxides, thereby mitigating significant sources of microwave loss.
Surface suboxides on germanium appear almost immediately after this surface treatment~\cite{sahari_native_2011} and could also constitute a major loss mechanism, limiting the microwave performance of devices on germanium even after HF treatment.

\sfref{fig:material}{c,d} show the power dependence of the microwave loss of Ge, HF-treated Ge, QW etched, and QW HF-treated.
We note that the power dependence of Ge is relatively flat compared to ``QW etched'', which does not saturate at $\langle n\rangle \approx 1$.
This might be due to a relatively low concentration of two-level systems (TLSs) intrinsically present in the epi-ready commercial Ge substrate, whereas the strain-relaxing composition grading of SiGe, different interfaces within the heterostructure, and/or the etching used to prepare the SiGe virtual layer may result in a higher concentration of TLSs.
We remark that HF treatment of both the Ge and Ge/SiGe QW heterostructures reduces $\delta^\text{LP}$ by about a factor of 2 and reduces $\delta^\text{HP}$ by an even larger factor.
This suggests that both target materials host significant power-independent losses, such as accumulated carriers that conduct at low temperature, and that this source of loss is reduced by HF and may reveal a larger relative TLS contribution after a power-independent background is reduced. 
This could occur directly on the surface itself or within a finite subsurface depth that is modified by the HF.

\subsection{Superconductors: PtSiGe and Nb-coated PtSiGe synthesis and materials characterization}

\begin{figure*}[ht]
\centering
\includegraphics[width = \textwidth]{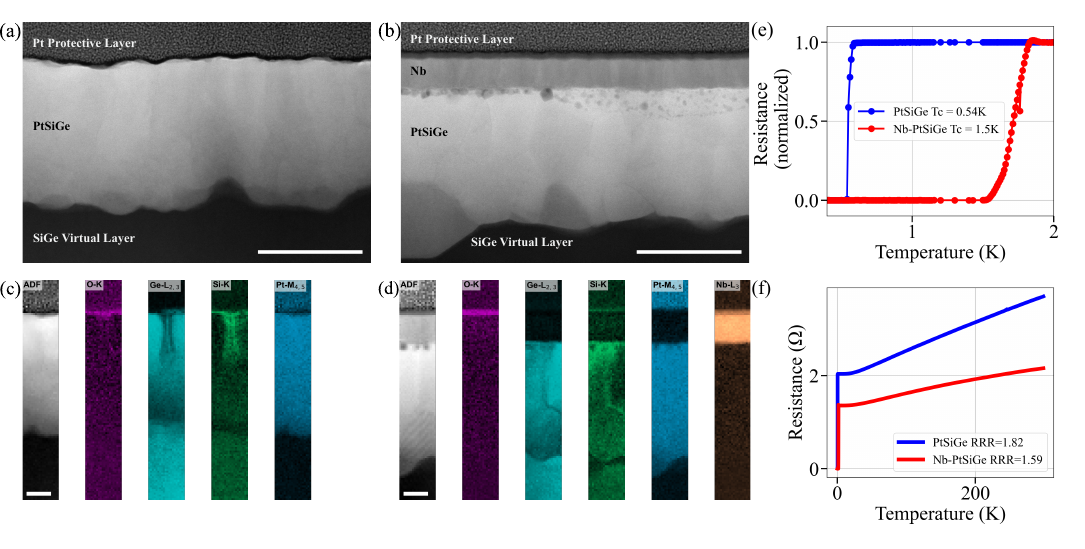}
\caption{\textbf{Scanning transmission electron microscopy (STEM) images of PtSiGe and Nb-coated PtSiGe (Nb-PtSiGe).} Low-magnification STEM ADF images of the (a) PtSiGe and (b) Nb-PtSiGe films.
Both films were kept under ambient conditions for 2 months prior to STEM imaging.
The scale bar is \SI{100}{\nano\meter}.
(c), (d) Reference ADF images and EELS maps of the $\mathrm{O\text{-}K}$,  $\mathrm{Ge\text{-}L_{2,3}}$,  $\mathrm{Si\text{-}K}$,  $\mathrm{Pt\text{-}M_{4,5}}$, $\mathrm{Nb\text{-}L_{3}}$ edges for the PtSiGe and Nb-PtSiGe samples, respectively. 
The scale bar is \SI{25}{\nano\meter}. 
(e) Resistance as a function of temperature for PtSiGe and Nb-PtSiGe films near the superconductor transition temperature.
(f) Residual resistivity ratio for PtSiGe and Nb-PtSiGe.
}
\label{fig:TEM} 
\end{figure*}

We now turn to the superconducting film PtSiGe, which has been found to induce a hard proximity gap when contacting 2DHGs in Ge/SiGe heterostructures in transport experiments~\cite{Tosato2023}.
Despite their success in hybrid devices for transport experiments, their suitability for microwave applications has not yet been reported.
PtSiGe is formed by depositing Pt onto a Ge/SiGe heterostructure followed by a rapid thermal annealing step.

PtSiGe can also be integrated into a heterostructure with a superconducting material exhibiting excellent microwave performance, reducing the amount of current flowing in PtSiGe and alleviating microwave losses.
Nb is widely used in conventional superconducting qubits due to its relatively low microwave losses, robust superconducting properties, and straightforward thin-film deposition~\cite{olszewski_low_2025, tuokkola_methods_2025, Bal2024}.
Motivated by these considerations, we fabricated Nb-coated PtSiGe (hereafter referred to as ``Nb-PtSiGe'') films to evaluate how Nb overcoating influences the microwave response and kinetic inductance of the superconductor.
This was done by depositing an Nb/Pt bilayer, followed by a rapid thermal annealing step.
Fabrication details for these films are presented in \aref{app:fab}.

We investigated the morphology and elemental distribution of PtSiGe and Nb-PtSiGe films using scanning transmission electron microscopy and electron energy loss spectroscopy (STEM-EELS).
\sfref{fig:TEM}{a} shows an annular dark-field (ADF) image of the \SI[separate-uncertainty=true]{138 \pm 7}{\nano\meter}-thick PtSiGe layer\footnote{Quoted uncertainty corresponds to one standard deviation.}.
No distinct layer corresponding to the Ge QW is observed, suggesting that it has been fully incorporated into the PtSiGe during formation.
This shows that PtSiGe formation reaches the depth of the buried QWs, including the \SI{66}{\nano\meter}-deep QW in the heterostructure studied here.

For the Nb-PtSiGe film shown in \sfref{fig:TEM}{b}, we observe a \SI[separate-uncertainty=true]{37 \pm 3}{\nano\meter}-thick Nb layer on top of a \SI[separate-uncertainty=true]{128 \pm 11}{\nano\meter}-thick PtSiGe layer. 
We note that the Nb coating reduces the surface roughness from \SI{4.3}{\nano\meter} for the PtSiGe film to \SI{2.4}{\nano\meter} for Nb-PtSiGe.

In both films, compositional mapping (\sfref{fig:TEM}{c,d}) and contrast changes in the ADF images indicate the formation and phase segregation of platinum germanide and silicide species. The EELS maps show anticorrelated Ge-$L_{2,3}$ and Si-$K$ signals within the PtSiGe layer, while the Pt distribution is uniform.  
Oxygen is found at the metal-air interface of both films, forming surface oxides with Si and Ge, and is largely absent in the PtSiGe layers.
The diffusion behavior of O at the metal-air interface also differs between the films. 
In the uncoated sample, O is found in a thin layer at the interface and diffused slightly downward through a Si-rich channel. 
At the metal-air interface of the coated film, O is found in a thicker niobium oxide layer and diffused throughout the Nb coating. 
Additionally, mapping the Ar-$L_{2,3}$ peak, we confirm that the dark spots between the Nb and PtSiGe layer in the ADF image are argon bubble formations; the source of Ar is likely from the sputtering process, which could be mitigated by using a Kr-based sputtering process, which reduces the noble gas impurities from the sputtering process~\cite{petrov_comparison_1993,paturaud_influence_1996,olszewski_krypton_2026}. 

We measured the resistance of these films as a function of temperature, shown in \sfref{fig:TEM}{e}.
We extracted a critical temperature of $T_\text{C}^\text{DC} = \SI{0.54}{\kelvin}$ for PtSiGe, consistent with recent reports~\cite{Tosato2023, li_germanide_2026}. 
For Nb-PtSiGe, we observed a higher critical temperature of $T_\text{C}^\text{DC} = \SI{1.5}{\kelvin}$. 
While this value remains below that of elemental Nb, it is higher than that of bare PtSiGe.
Given that the PtSiGe layer is more than three times thicker than the Nb overlayer, the bilayer transition temperature is expected to remain significantly influenced by PtSiGe. 
This is qualitatively consistent with the generalized proximity-effect model of Ref.~\cite{Brammertz2002}, which predicts the transition temperature of a superconducting bilayer depends on the layer thicknesses and interface properties, and demonstrated that increasing the thickness of the lower-$T_\text{C}$ layer shifts the common bilayer transition temperature toward that of the lower-$T_\text{C}$ material.

\subsection{Superconductors: microwave losses}\label{sec:sc_film_loss}

\begin{figure*}[htb]
\centering
\includegraphics[width = \textwidth]{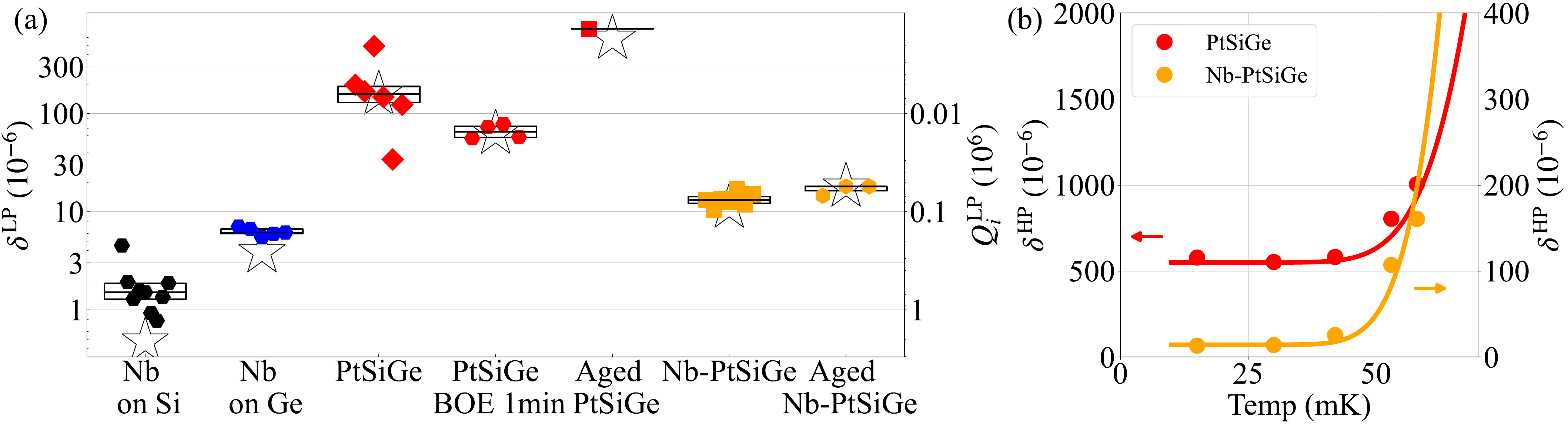}
\caption{ 
\textbf{Superconducting thin film loss.} 
(a) Box plot of average low-power (all $\braket{n}<10$) intrinsic loss $\delta^\text{LP}$ (dots) and average high-power (defined here as the average of the largest two photon number points from all resonators with $\braket{n} < 10^7$) intrinsic loss $\delta^\text{HP}$ (stars) for superconducting thin films measured by the flip-chip-based sensor.
The PtSiGe and Nb-coated PtSiGe (Nb-PtSiGe) samples are remeasured after 2 months of aging.
(b) Temperature dependence of high power $\delta^\text{HP}$ for PtSiGe and Nb-PtSiGe. Solid points indicate the measured values, and solid lines represent the Mattis-Bardeen fits. 
We extract RF critical temperatures of $T_\text{C}^\text{RF}=0.30~\text{K}$ for PtSiGe and $T_\text{C}^\text{RF}=0.33~\text{K}$ for Nb-PtSiGe, assuming $\Delta_0 = 1.764k_\text{B}T_\text{C}$ where $\Delta_0$ is the superconducting gap.
}
\label{fig:metal} 
\end{figure*}

Building on the structural and transport characterization above, we use the flip-chip sensor to probe the microwave response of these superconducting films.
\sfref{fig:metal}{a} shows the measured microwave losses of the flip-chip-based sensor for different target superconducting films.
To calibrate the setup, we sensed a \SI{60}{\nano\meter} Nb film on a Si substrate, obtaining a low-power ($\delta^\text{LP}\text{sensor} = (1.5 \pm 1.0)\times10^{-6}$) and high-power ($\delta^\text{HP}\text{sensor} = (0.5 \pm 0.3)\times10^{-6}$) loss of the sensing resonator.
We then measured a \SI{60}{\nano\meter} Nb film on a bare Ge substrate, obtaining $\delta^\text{LP} = (6.1 \pm 0.6)\times10^{-6}$ and $\delta^\text{HP} = (3.9 \pm 0.7)\times10^{-6}$.
The higher loss observed for Nb on Ge suggests increased surface oxide and/or conductor loss, likely arising from differences in the metal–substrate interface or the growth conditions for the niobium layer, which will be investigated more in future work.

For the as-grown PtSiGe film, we measure $\delta^\text{LP} = (158 \pm 141)\times10^{-6}$, which increases to $(736 \pm 112)\times10^{-6}$ after two months of ambient exposure. 
Motivated by the presence of silicide and germanide oxides observed in \sfref{fig:TEM}{c}, as well as the strong increase in loss upon aging, we performed a buffered oxide etch (BOE) treatment on a newly fabricated PtSiGe film. 
X-ray photoelectron spectroscopy (XPS) measurements (\aref{app:xps}) confirmed that this treatment removes a portion of the surface oxides.
Following BOE treatment, we measure a lower loss of $\delta^\text{LP} = (65.3 \pm 9.7)\times10^{-6}$. 

In contrast, Nb-PtSiGe exhibits a much lower loss of $\delta^\text{LP} = (13.2 \pm 2.0)\times10^{-6}$, which increases to $\delta^\text{LP} = (18.1 \pm 1.7)\times10^{-6}$ after two months of ambient exposure. 
This reduced sensitivity to aging is consistent with the relatively slow growth of the native surface oxide of Nb~\cite{verjauw_investigation_2021}.
The high microwave loss of PtSiGe is potentially a combined effect of conductor loss and surface oxide loss. 
The observed aging suggests that PtSiGe native oxides may be less stable under ambient conditions than the Nb native oxide.
An estimation of bounds on the RF sheet resistance, calculated using measured microwave loss, can be found in \aref{app:conductor}.

We further measure the high-power loss $\delta^\text{HP}$ of both PtSiGe and Nb-PtSiGe as a function of temperature, shown in \sfref{fig:metal}{b}.~\footnote{Note that we expect the experiment to be sensitive to the PtSiGe layer in Nb-PtSiGe because the penetration depth of Nb is comparable to the thickness of the Nb layer.}
By fitting the temperature dependence using Mattis–Bardeen theory~\cite{mattis_theory_1958}, we extract RF critical temperatures of $T_\text{C}^\text{RF} = 0.30~\text{K}$ for PtSiGe and $T_\text{C}^\text{RF} = 0.33~\text{K}$ for Nb-PtSiGe. 
The fitting parameters are $A_\text{QP}=0.1$, $1/Q_\text{other}=550\times 10^{-6}$, $f_\text{r}=\SI{7.3}{\giga\hertz}$ for PtSiGe and $A_\text{QP}=0.1$, $1/Q_\text{other}=14.3\times 10^{-6}$, $f_\text{r}=\SI{4.7}{\giga\hertz}$ for Nb-PtSiGe, assuming temperature-dependent loss is dominated by QP loss, following the model from~\cite{crowley_disentangling_2023}.
These values are significantly lower than the corresponding $T_\text{C}^\text{DC}$ extracted from transport measurements.
The disagreement in $T_C$ values for PtSiGe measured by transport and resonators is consistent with gap inhomogeneity or another broadened/subgap resonance in the superconducting gap of PtSiGe. 
The discrepancy for Nb-PtSiGe further indicates that the \SI{\approx 40}{\nano\meter} Nb layer is consistent with incomplete proximitization of low-gap regions, although gap-broadening mechanisms are not excluded. 

Combining these results with the loss decomposition above, we find that Nb coating suppresses the total PtSiGe-associated loss, as well as the aging-induced increase in dissipation. 
However, an attempt to fabricate an Nb-PtSiGe coplanar strip resonator did not yield a measurable resonance.
This resonator was designed with a coupling quality factor of about 3000, which suggests an internal quality factor more than 10$\times$ lower.
Possible loss channels include the metal-substrate (MS) interface and resonator sidewalls that expose a partial PtSiGe metal-air interface.
This metal-air sidewall is expected to have a surface participation ratio of around $46\times 10^{-6}$; combined with our estimated bound for the loss tangent of the oxidized PtSiGe surface, this gives an estimated low-power resonator internal quality factor of around $500$.
This is insufficient by itself to account for the undetectable resonance peak.
The metal-substrate interface is one possible additional loss channel, which is not probed by the flip-chip method because the electric fields are screened.
The thermal annealing approach may introduce chemical gradients of dilute Pt into the SiGe and Ge that introduce microwave losses at the MS interface; this correlation was previously observed for Pt on Si~\cite{szypryt_ultraviolet_2015,szypryt_high_2016}.

\subsection{Superconductors: kinetic inductance}

The kinetic inductance of a target superconducting film modifies the resonance frequency of the sensing resonator.
This enables extraction of the film kinetic inductance per square $L_K^\square$ using the flip-chip sensor over a range of $0-\SI{10}{\pico\henry}/\square$, with a sensitivity corresponding to approximately a 1\% resonance-frequency shift per $\SI{1}{\pico\henry}/\square$.

As a verification and calibration step, we first measured a \SI{50}{\nano\meter} $\delta$-NbN film grown by molecular beam epitaxy.
We extract $L_K^\square$ of \SI[separate-uncertainty=true]{6.9\pm1.1}{\pico\henry}$/\square$, in reasonable agreement with the calculated value of $\SI{7.7}{\pico\henry}/\square$ estimated from transport data.
We determine kinetic inductances of $L_K^\square=\SI[separate-uncertainty=true]{3.8\pm0.2}{\pico\henry}/\square$ and $L_K^\square = \SI[separate-uncertainty=true]{3.0\pm0.2}{\pico\henry}/\square$ from the flip-chip method for PtSiGe and Nb-PtSiGe, respectively.
These kinetic inductance values are directly relevant for circuit design, as they determine impedance, nonlinearity, and coupling strengths in superconducting quantum circuits.
Further details on the extraction procedure and modeling of the kinetic inductance are provided in \aref{app:KI}.

\section{Discussion}
We demonstrated a fabrication-free method to measure target-associated effective microwave losses in quantum materials.
By employing SU-8--based pillars, we eliminate the need for indium bumps traditionally used in flip-chip assemblies, allowing the method to be extended to geometries that do not require galvanic contact.
We applied this approach to a range of germanium-based materials, including epitaxial heterostructures and solid-state-reaction-based superconductors.
In the future, the cross-sectional geometry of the resonator or the resonator--material distance could be varied to more directly separate different sources of loss.

In the heterostructure used in this study and in Ref.~\cite{Ruggiero2026}, the virtual layer and quantum well are grown directly on a Ge substrate.
This eliminates the need to introduce a heavily plastically strain-relaxed Ge starting layer, as in reverse-composition-grading approaches on Si substrates~\cite{Nigro_loss_QW_2024, Palma_preprint_2025}.
Due to the 4.7\% lattice mismatch between Si and Ge, the interface between such a Ge starting layer and a Si substrate induces an abrupt plastic relaxation via the inclusion of crystalline defects, in particular dislocations~\cite{sammak_shallow_2019}.
The architecture reported here may be advantageous for microwave resonators because electric fields penetrate several micrometers into the substrate.
Notably, lower microwave losses have also been observed in heterostructures that avoid such a strain-relaxed Ge starting layer through forward composition grading~\cite{valentini_parity_2024}.
This is further corroborated by our measurements, which show that the strain-engineering scheme used here exhibits microwave losses comparable to those of the bare epi-ready commercial Ge substrate.
We validated this conclusion by fabricating a Nb coplanar-strip resonator on an etched SiGe heterostructure, obtaining loss values consistent with the flip-chip measurements and confirming that no large additional fabrication-associated loss is resolved.
In the future, we can apply our technique to the other heterostructures noted here for a consistent comparison without the uncertainties introduced by different fabrication processes.

We found that residual conduction in the Ge/SiGe QW heterostructure at cryogenic temperatures can be a significant source of microwave loss.
Surface treatments can significantly change the residual conduction and reduce surface oxides, leading to improved microwave performance beyond the studied transport properties~\cite{Sangwan2025}.
Further studies can help identify surface preparations that may be suitable for a variety of Ge/SiGe heterostructures that are now grown by different groups~\cite{Nigro2025, stehouwer_exploiting_2025, sammak_shallow_2019, Oezkent_preprint_2026, Daoust_nuclearfree_2026, myronov_hole_2025}.

We also examined PtSiGe, a thin-film superconductor formed through a solid-state reaction process that is known to form highly transparent contacts to Ge quantum wells.
We found that its apparent microwave losses are relatively high, restricting its suitability as a standalone superconductor for qubit circuitry.
Coating PtSiGe with Nb substantially reduced these losses in our configuration, when Nb is able to screen electric fields from reaching the PtSiGe.
Additionally, the temperature dependence is consistent with an incomplete proximity effect or another broadened-gap mechanism.
Future work can focus on alternative reaction-based metal--SiGe thin films~\cite{li_germanide_2026}, planar-tunneling-based proximity methods~\cite{valentini_parity_2024, Ruggiero2026}, or other bilayer superconductors to achieve lower-loss superconducting layers.
These findings establish key material constraints for hybrid superconducting--semiconductor quantum devices, particularly highlighting the limitations of solid-state-reaction-based superconductors such as PtSiGe for low-loss microwave applications.

Overall, these results inform the choice of superconductors and heterostructures for low-loss microwave circuit design.
More broadly, the measurement approach demonstrated here provides a general platform for rapid screening and optimization of emerging materials for quantum technologies, helping to constrain intrinsic and extrinsic loss mechanisms. 

\section*{Author Contributions}
H. L. and K. A. designed the device, processed the SiGe-related materials, acquired the data, and analyzed the data under the guidance of V. F..
H. L., M. W. O., L. K., and A. I. developed the fabrication process and fabricated the sensors.
X. W. and H. L. conducted the simulation under the guidance of V. F. and P. L. M..
P. D. deposited the heterostructure under the guidance of D. B..
D. T. collected and analyzed the STEM images under the guidance of D. A. M..
M. W. O. deposited the Ta film for the sensor.
A. I. deposited the NbN film and collected the transport data under the guidance of D. J..
L. K. conducted the XPS analysis.
V. F. conceived the method and established the collaboration with D. B.. 
H. L., K. A., and V. F. wrote the manuscript with input from all authors.

\section*{Data and code availability}
All data generated and code used in this work are available at: \textcolor{black}10.5281/zenodo.22964967.

\section{AI usage disclosure}
We used OpenAI Codex with GPT-5.6 and GPT-6 models to help generating plotting codes and proofread the text.

\begin{acknowledgments}

We acknowledge assistance from Corey Rae McRae on the microwave packaging and resonator design, discussions with Luojia Zhang regarding the SU-8 recipe, discussions with Doug Bennet regarding modeling dielectric loss in HFSS, and discussions with Tathagata Banerjee regarding XPS fitting and analysis.

The flip chip method for sensing microwave properties was developed under the Air Force Office of Scientific Research under award number FA9550-23-1-0688. Any opinions, findings, and conclusions or recommendations expressed in this material are those of the author(s) and do not necessarily reflect the views of the United States Air Force.
We acknowledge support by the Army Research Office under Grant Number W911NF-22-1-0053 for assessing Ge-based materials.
This research is supported by The Kavli Foundation. 
This research is part of the Munich Quantum Valley, which is supported by the Bavarian state government with funds from the Hightech Agenda Bavaria.
This work was performed in part at the Cornell NanoScale Facility, a member of the National Nanotechnology Coordinated Infrastructure (NNCI), which is supported by the National Science Foundation (Grant NNCI-2025233).
This work made use of the Cornell Center for Materials Research shared instrumentation facility.
This work made use of the Meehl cryostat donated by David W. Meehl in memory of his father James R. Meehl and supported by the Cornell College of Engineering.
\end{acknowledgments}

\section*{Competing Interests}
The authors declare no competing interests.

\appendix

\section{Heterostructure growth}\label{app:heterostructure}

In this study, we used commercial, epi-ready Ge substrates with a thickness of \SI{475 \pm 25}{\micro\meter} and a room-temperature resistivity of \SI{>30}{\ohm\centi\meter}.
The growth of the undoped, rectangular QW $\text{Ge}_{0.75}\text{Si}_{0.25}\text{/Ge/Ge}_{0.75}\text{Si}_{0.25}$ heterostructure---hereafter denoted the \textit{Ge/GeSi heterostructure}---was performed in a solid-source molecular-beam-epitaxy system operating at a base pressure below \SI{5e-10}{\milli\bar}.
The native Ge oxide of the substrate was removed by thermal deoxidation, followed by growth of a \SI{180}{\nano\meter} Ge initiation buffer layer to prepare an atomically flat surface.
To engineer gradual strain relaxation and provide a virtual substrate for strain-relaxed $\text{Ge}_{0.75}\text{Si}_{0.25}$, a linearly composition-graded $\text{Ge}_x\text{Si}_{1-x}$ layer was grown, in which the Ge content $x$ was reduced from 100\% to 75\% over \SI{520}{\nano\meter}.
The undoped Ge/GeSi heterostructure consists of a \SI{370}{\nano\meter} constant-composition $\text{Ge}_{0.75}\text{Si}_{0.25}$ layer, serving as the lower QW potential barrier; a \SI{15}{\nano\meter} fully strained Ge QW; and a \SI{66}{\nano\meter} $\text{Ge}_{0.75}\text{Si}_{0.25}$ top spacer, acting as the upper QW potential barrier.
To protect the SiGe from oxidation in ambient air, we added a \SI{1.5}{\nano\meter} crystalline Si cap on top of the spacer.

\section{Device fabrication}\label{app:fab}

\subsection{Sensor chip}

\subsubsection{Base resonators}

A \SI{100}{\nano\meter} tantalum base layer with a \SI{5}{\nano\meter} Nb seed layer was deposited on a 4-inch high-resistivity Si wafer.
We used the same fabrication and post-fabrication cleaning recipes as Ref.~\cite{olszewski_krypton_2026}.
The base resonator chip was covered with S1813 resist (and later stripped before SU-8 lithography) to protect the base layer from contamination.

\subsubsection{SU-8 pillar}

The processed base resonator chip was covered by SU-8 2005 photoresist spun at 4000 rpm.
The chip was placed on a clean glass slide for $\SI{5}{\minute}$ to let the SU-8 resist flow, forming a more uniform surface~\cite{norri_flips_2024,ezkerra_fabrication_2007}.
This was followed by a baking step at \SI{65}{\celsius} for \SI{90}{\second} and \SI{95}{\celsius} for \SI{90}{\second}.
The devices were patterned using a Heidelberg MLA 150 Maskless Aligner and a dose of 4000 mJ/cm$^2$. Before development, the chip was baked at \SI{65}{\celsius} for \SI{90}{\second} and \SI{95}{\celsius} for \SI{90}{\second}.
The chip was developed in SU-8 thinner for \SI{60}{\second} followed by a \SI{30}{\second} IPA rinse.
If any residues from SU-8 were observed, an additional \SI{5}{\second} SU-8 thinner development and \SI{30}{\second} IPA rinse was performed.
After the development, the chip was soaked in AZ300T bath for \SI{9}{\minute}, IPA bath for \SI{5}{\minute}, and DI water bath for \SI{3}{\minute} for a deeper clean. 

The chip was then baked at \SI{90}{\celsius} for \SI{15}{\minute} to further harden the SU-8 polymer.
The baking time was chosen to be short as SU-8 film stress can increase drastically after long baking \cite{hammacher_stress_2008}.
The height of the SU-8 pillar was measured with a profilometer for quality control.

\subsection{PtSiGe and Nb-coated PtSiGe films}
The heterostructure wafer was first diced into smaller samples. The wafer was coated with photoresist to protect it during dicing.
After dicing, the samples were cleaned in 1165 stripper, followed by rinses in IPA and DI water.
A 15 s buffered oxide etch was performed, followed by a DI water rinse.
The samples were immediately transferred to the sputtering system.
A \SI{60}{\nano\meter} Pt layer (PtSiGe) or a \SI{60}{\nano\meter} Pt/$\SI{\approx40}{\nano\meter}$ Nb bilayer (Nb-PtSiGe) was sputtered.
After deposition, the samples were annealed in a rapid thermal annealer for 9 min at \SI{400}{\celsius} in an Ar-rich environment to form PtSiGe and Nb-PtSiGe alloys via a solid-state reaction.

\subsection{SiGe virtual layer sample}\label{app:virtualSiGe}
The diced heterostructure chip was first cleaned in 1165 stripper followed by rinse in DI water.
The top \SI{66}{\nano\metre} Si$_{0.25}$Ge$_{0.75}$ and \SI{15}{\nano\metre} Ge quantum well were etched in an Oxford PlasmaLab 80+ RIE system with SF$_6$, CHF$_3$ and O$_2$ gases in a ratio of 26:26:17 sccm with a RIE power of 50 W at 5 mTorr. 

\subsection{Coplanar strip resonator on SiGe virtual layer}\label{app:cpw}
For the resonator, the sample was first prepared as described in \aref{app:virtualSiGe}.
After the etch, a \SI{15}{\second} buffered oxide etch was performed followed by rinse in DI water.
A \SI{60}{\nano\metre} Nb film was sputtered onto the chip. The chip was spun with S1813 resist. The coplanar strip resonator was patterned using the Heidelberg MLA 150 Maskless Aligner.
The chip was developed in MIF726 for \SI{1}{\minute} and rinsed in DI water for \SI{30}{\second}.
The chip was subjected to a light etch with a mixture of BCl$_{3}$, Cl$_2$, and Ar gas in a 2:30:5 sccm ratio with a RIE/ICP power of 26/800 W at 13 mTorr.
This was followed by a primary etch with a mixture of   BCl$_{3}$, Cl$_2$, and Ar gas in a 30:20:5 sccm ratio with a RIE/ICP power of 12/800 W at 7 mTorr. 
The chip was finally cleaned in AZ300T followed by rinse in IPA and then DI water.

\subsection{HF treatment of Ge/SiGe heterostructure}
The Ge bare substrate and the SiGe virtual layer substrate were treated in 5\% HF for \SI{30}{\second}, followed by a DI water rinse.
The HF-treated samples were transferred to the flip-chip sensor and loaded into a cryostat roughly within one hour.

\section{Conductance of as-grown Ge/SiGe heterostructure}\label{app:dcdata}
We performed electrical characterization of the presented QW heterostructure at \SI{1.6}{\kelvin} in a Hall bar geometry using a standard low-frequency lock-in technique, comparing as-grown samples with HF surface-treated samples (\SI{30}{\second} dip in 5\% HF).
While as-grown samples were consistently found to be conductive, even at cryogenic temperatures, the HF surface-treated samples were consistently found to be nonconductive and remained nonconductive for at least six months without further treatment while being stored in ambient air.

\section{Device design and resonance fitting}
\subsection{Resonator design}
The sensing resonator design is adapted from a standard NIST design \cite{kopas_simple_2022}.
The NIST design has eight $\lambda/4$ resonators between \SI{4}{\giga\hertz} and \SI{8}{\giga\hertz}, with a gap-metal-gap ratio of \SI{3}{\micro\meter}-\SI{6}{\micro\meter}-\SI{3}{\micro\meter}.

In this work, the sensing resonator has an identical layout and resonator length, but with a gap-metal-gap of \SI{12}{\micro\meter}-\SI{6}{\micro\meter}-\SI{12}{\micro\meter}. 
The increased gap size is meant for maintaining \SI{50}{\ohm} impedance when a metallic target top chip is placed and increases the participation ratio to the target top chip.

\subsection{Resonance fitting}
We extract complex coupling quality factor $\hat{Q_{c}}$, internal quality factor $Q_{i}$, and center frequency $f_r$ using the diameter correction method (DCM)~\cite{mcrae_materials_2020,khalil_analysis_2012}:
\begin{equation}
    \label{eq:S21-fit}
    S_{21}(f) = 1-\frac{Q/\hat{Q_{c}}}{1+2iQ\frac{f-f_{r}}{f_{r}}}.
\end{equation}

The average photon number was calculated using~\cite{gao_physics_2008,bruno_reducing_2015}:
\begin{equation}
    \langle n\rangle=\frac{2}{\hbar\omega_{0}^{2}}\frac{Z_{0}}{Z_{r}}\frac{Q^{2}}{Q_{c}}P_\mathrm{app},
\end{equation}
where $\omega_0$ is the angular resonance frequency, $Z_0=\SI{50}{\ohm}$ is the characteristic impedance of the microwave environment the resonator couples to, $Z_r$ is the characteristic impedance of the resonator, $Q$ is the total quality factor, $Q_c$ is the coupling quality factor,  and $P_\text{app}$ is the RF power at the sample.
We adapted an open-source resonator-fitting code from NIST Boulder into our fitting procedure~\cite{scresonators}.

\section{Sensing kinetic inductance}\label{app:KI}
\begin{figure}
    \centering
    \includegraphics[width=1\linewidth]{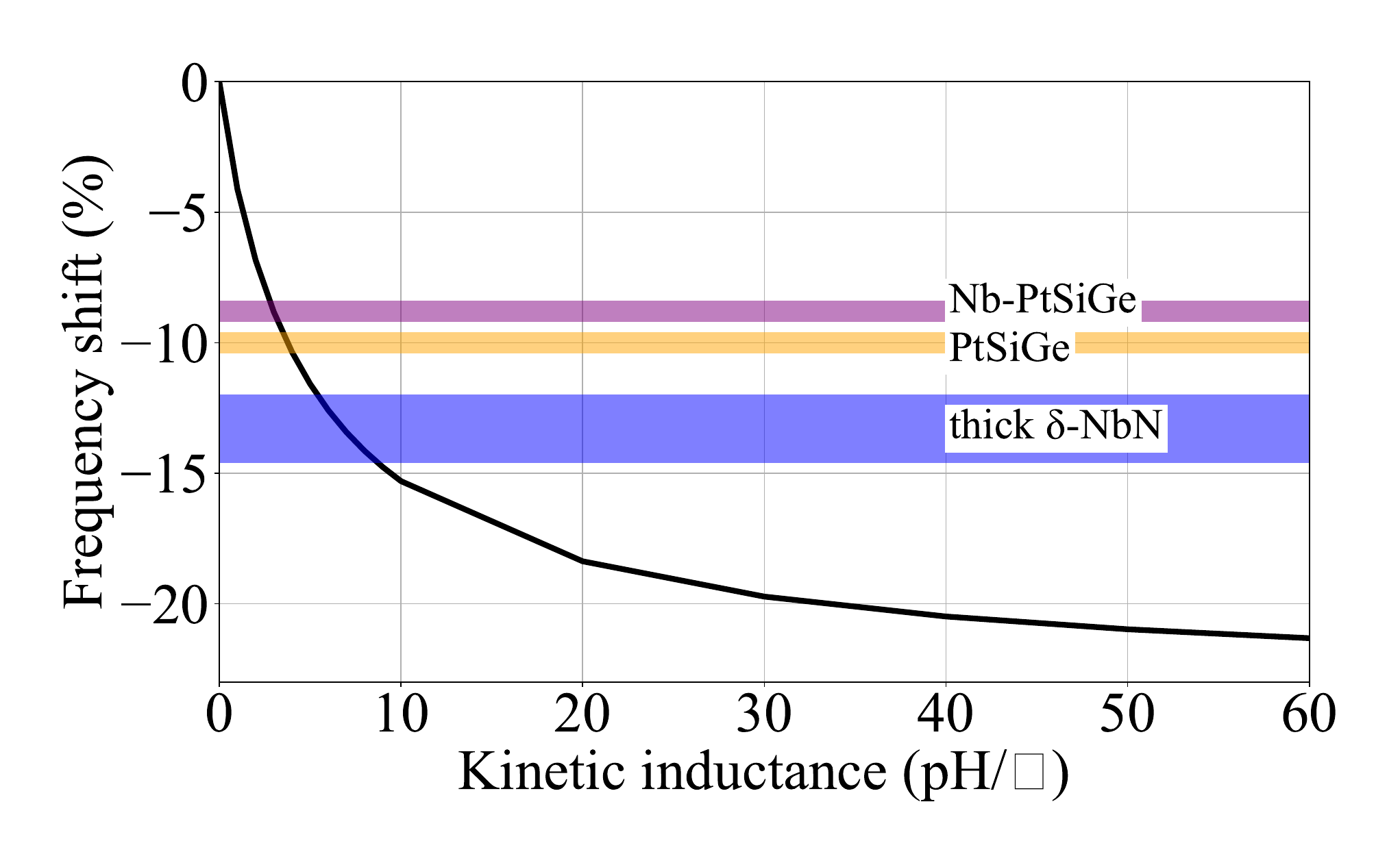}
    \caption{\textbf{Kinetic inductance extraction.} 
    Black line: predicted frequency shift at target films of different kinetic inductances.
    Colored bands: frequency shifts observed from experiments. 
    Nb on Si target film is used as a baseline resonance frequency for low-kinetic inductance.
    The intersection of each colored band with the simulated curve is projected onto the x-axis to obtain the corresponding kinetic-inductance range.
    }
    \label{sfig:KI}
\end{figure}

Kinetic inductance of the superconducting film on the target top chip will induce a downward shift of the resonance frequency of the sensor. 
\fref{sfig:KI} shows the frequency shift of the sensor coupled to target superconducting films of different kinetic inductance simulated by Sonnet Suites\textsuperscript{TM}.
The predicted kinetic inductance per square of $\delta$-NbN, PtSiGe, and Nb-PtSiGe target films are $\SI[separate-uncertainty=true]{6.9\pm1.1}{\pico\henry}/\square$, $\SI[separate-uncertainty=true]{3.8\pm0.2}{\pico\henry}/\square$, and $\SI[separate-uncertainty=true]{3.0\pm0.2}{\pico\henry}/\square$,  respectively.

The kinetic inductances of the $\delta$-NbN, PtSiGe, and Nb-PtSiGe films are estimated using~\cite{mattis_theory_1958,tinkham_introduction_2004}:
\begin{equation}
L^{\square}_K=\frac{\hbar R^{\square}}{\pi \Delta},
\end{equation}
where $\hbar$ is the reduced Planck constant, $R^{\square}$ is the resistance per square measured close to superconducting transition temperature, and $\Delta$ is the superconducting gap.
$R^{\square}=55,2.4,\SI{6.1}{\ohm}/\square$, $\Delta=1.5,0.082,\SI{0.23}{\milli\eV}$, and $L^{\square}_K=7.7,6.1, \SI{5.6}{\pico\henry}/\square$ for $\delta$-NbN film, PtSiGe, and Nb-PtSiGe respectively.
We remark that $R^{\square}$ of PtSiGe is estimated via van der Pauw measurement 10 months after growth at room temperature and normalized by the RRR of PtSiGe, resulting in potential systematic changes.
$R^{\square}$ of Nb-PtSiGe is extracted via collinear four-point probe measurement on an irregular piece (\sfref{fig:TEM}{f}).

We note that larger kinetic inductance values remain detectable; however, the associated uncertainty increases due to the reduced linearity of the frequency shift at larger kinetic inductance.

\section{Calculation of participation ratios}

Resonator loss $\delta$ can be expressed as~\cite{garcia_disentangling_2024}:
\begin{equation}
    \delta = \delta^{\rm diel} + \delta^{\rm cond},
\end{equation}
where $\delta^{\rm diel}$ is the dielectric loss and $\delta^{\rm cond}$ is the conductor loss.
The dielectric loss of our flip-chip sensor is determined by the energy participation ratio in and the loss tangent of each domain~\cite{Kosen_flip_2022}:
\begin{equation}\label{app:di_loss}
    \delta^{\rm diel} = \sum_i pr_{i} \tan\delta_{i}, 
\end{equation}
where $pr_{i}$ is the participation ratio and $\tan\delta_i$ is the loss tangent of each domain. 

In the following sections, we discuss the relationships between the measured flip-chip sensor microwave loss and the loss tangent of the dielectric material, the resistance per square, and the surface oxide loss tangent for the superconducting thin film. 

\subsection{Dielectric material loss simulation}\label{app:pr_mat}
\subsubsection{Simulation and obtained values}
\begin{figure}
    \centering
    \includegraphics[width=1\linewidth]{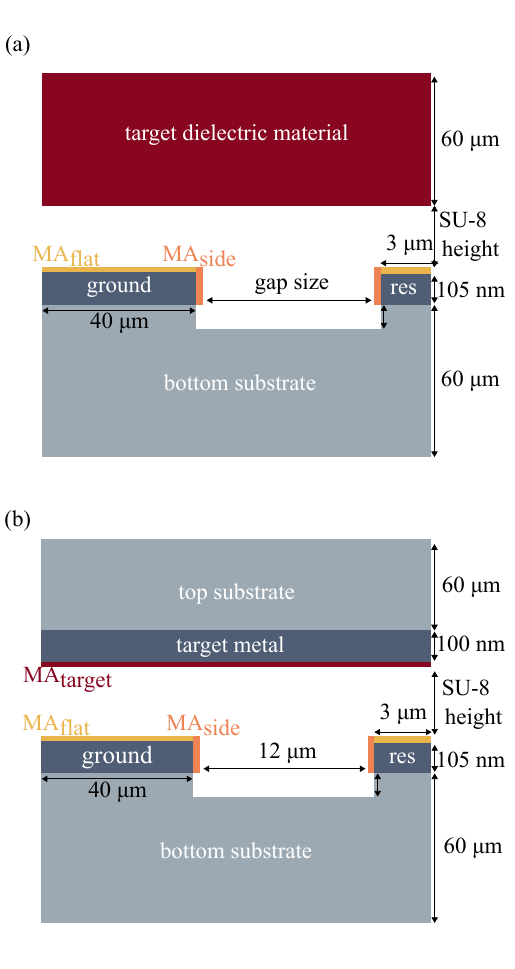}
    \caption{\textbf{Simulation setup for assessing microwave loss in flip-chip setup.}
    (a) Schematic cross-section to assess target dielectric material. We sweep the air gap and gap size to estimate the sensitivity of our device to the target material.  
    (b) Schematic of the flip-chip-based sensing scheme to extract surface loss of superconducting films. We sweep air gap or SU-8 height in simulation.  
}
    \label{sfig:bulk_dielectric}
\end{figure}

The loss tangent of the dielectric material is calculated by normalizing the additional microwave loss contributed from the target material over the energy participation ratio of the target material $pr_\text{material}$.
To obtain $pr_\text{material}$, we simulate its cross-section in Ansys Q2D Extractor, as shown in ~\sfref{sfig:bulk_dielectric}{a,b}. 
We set up a simulation including target material, vacuum, the flat top metal-air oxide of the sensing resonator ($\text{MA}_\text{flat}$), the side wall metal-air oxide of the sensing resonator ($\text{MA}_\text{side}$), and the bottom substrate.
The metal regions are assigned as perfect conductors: we define the resonator and the ground as the signal line and reference ground in Ansys Q2D, respectively.

The thicknesses of various domains can be found in \sfref{sfig:bulk_dielectric}{a}. 
Since the loss tangent only enters Eq.~\eqref{app:di_loss} as a scaling factor, and is typically a few orders of magnitude smaller than the permittivity $\epsilon_r$, we model the dielectric as lossless in the simulation without invalidating the comparison between iterated designs~\cite{von_dielectric_1958}. 

\begin{figure}
    \centering
    \includegraphics[width=0.95\linewidth]{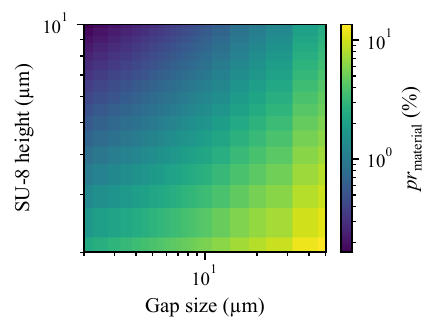}
    \caption{\textbf{Simulated target material participation ratio.}
    The color map shows the target material participation ratio as a function of air gap from \SI{2}{\micro\meter} to \SI{10}{\micro\meter} and gap size from \SI{2}{\micro\meter} to \SI{50}{\micro\meter}. A smaller air gap and a larger gap size lead to a higher participation ratio in the target material.
}
    \label{sfig:bulk_dielectric_cm}
\end{figure}

We vary the coplanar waveguide gap size from \SI{2}{\micro\meter} to \SI{50}{\micro\meter} and the SU-8 height from \SI{2}{\micro\meter} to \SI{10}{\micro\meter} in simulation to optimize $pr_\text{material}$. 
The simulation shows a wide range of tunability of $pr_\text{material}$, from 0.165\% at SU-8 height of \SI{10}{\micro\meter} and gap size of \SI{2}{\micro\meter} to 13.6\% at SU-8 height of \SI{2}{\micro\meter} and gap size of \SI{50}{\micro\meter} (\fref{sfig:bulk_dielectric_cm}). 
By adopting SU-8 pillars of different heights, the sensing resonator can efficiently detect additional loss contributed by the target material between $0.5\times 10^{-6}$ and $100\times 10^{-6}$, corresponding to a target material loss tangent detection range of $3.686\times 10^{-6}$ to $6.07\times 10^{-2}$, depending on the chosen distances and other design factors that tune $pr_\text{material}$. 
In this work, the SU-8 height is set to \SI{4}{\micro\meter} to target the losses relevant for Ge-related materials, as shown in ~\sfref{fig:material}{a}.

\subsubsection{Electric field participation depth profile}

\begin{figure}
    \centering
    \includegraphics[width=0.95\linewidth]{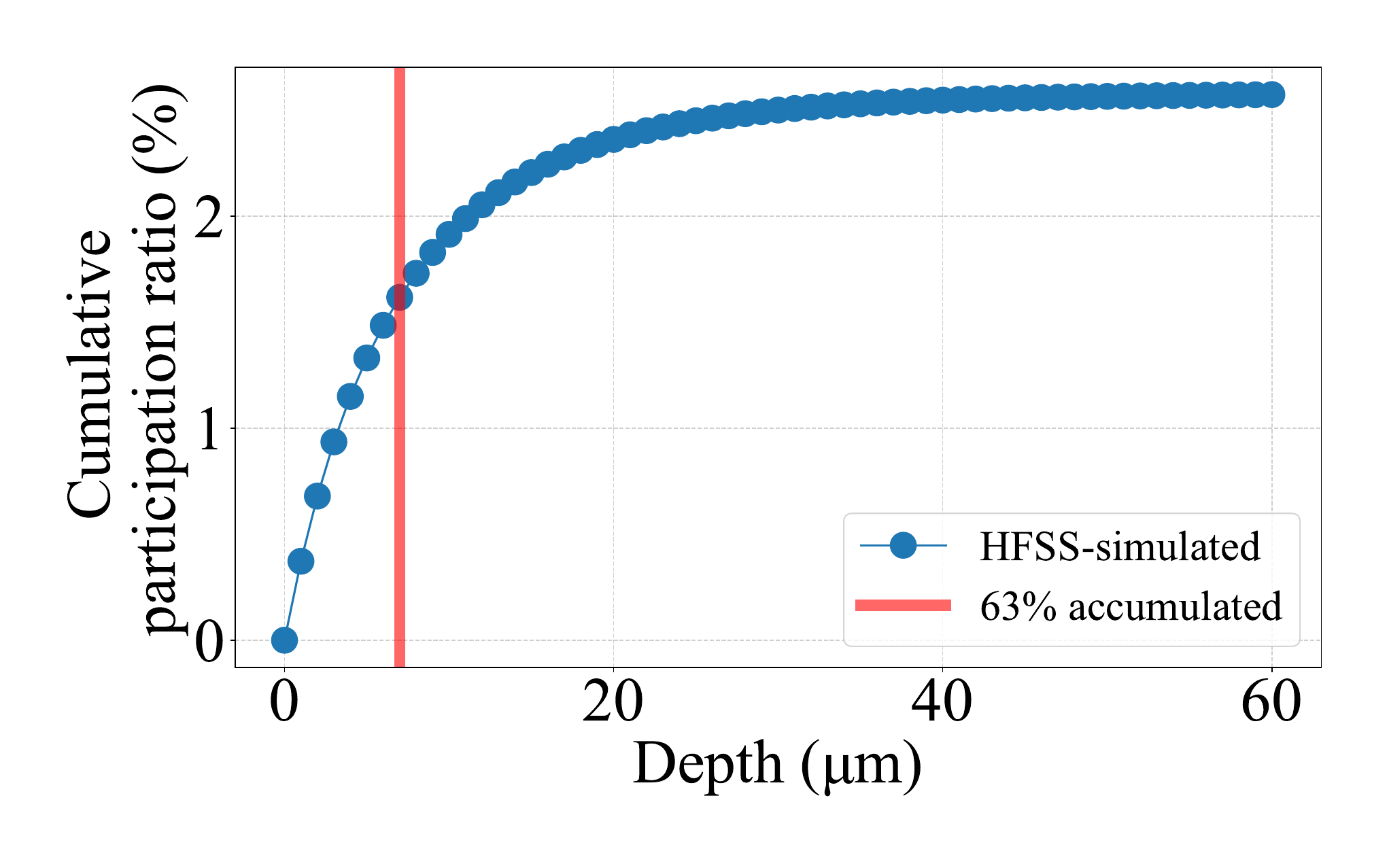}
    \caption{\textbf{Electric energy participation ratio depth profile for flip-chip sensor.}
    Cumulative energy participation ratio distribution in target dielectric material of $\epsilon_r=16.2$ (Ge) simulated via HFSS.
    Approximately 63\% ($1 - 1/e$) of the electric energy is stored within the first \SI{7}{\micro\meter}, establishing the characteristic decay length.
}
    \label{sfig:penetration}
\end{figure}

The electric field energy decays with depth within the target dielectric material, enabling detection of surface dielectric loss as well as high-loss-tangent interfaces/layers buried in the heterostructure.
The simulated cumulative electric field energy in the target dielectric material is shown in~\fref{sfig:penetration}.

The electric energy decay in the target dielectric material follows a multiexponential profile~\cite{simons_coplanar_2004}.
For germanium ($\epsilon_r=16.2$), approximately 63\% of the electric energy is concentrated within the first $\SI{7}{\micro\meter}$, corresponding to the characteristic decay length where $1-1/e \approx$ 63\%.

For the target material involved in this study, the roughly \SI{1.15}{\micro\meter} Ge-rich heterostructure that is grown upon the Ge substrate constitutes 18\% of the electric field energy inside the target material.

\subsubsection{Surface oxide microwave loss for superconducting thin film}\label{app:MA}
The PtSiGe films studied in this work have thick surface oxides, as shown in~\sfref{fig:TEM}{c}.
To assess the loss tangent of the metal-air surface oxide $\text{MA}_\text{target}$, we obtain its energy participation ratio using Ansys Q2D simulation.
We set up a simulation including the metal-air oxide of the target film ($\text{MA}_\text{target}$), the target film substrate, the vacuum, the flat top metal-air oxide of the sensing resonator ($\text{MA}_\text{flat}$), the side wall metal-air oxide of the sensing resonator ($\text{MA}_\text{side}$), and the substrate.
Details of the domains are illustrated in~\sfref{sfig:bulk_dielectric}{b}.
The relative permittivity $\epsilon_r$ of $\rm MA_{target}$,  target film substrate, vacuum, $\rm MA_{flat}$, $\rm MA_{side}$, and the bottom substrate are 10, 16.2, 1, 20, 20, and 11.45, respectively. 
All metal-air interfaces ($\rm MA$) are set to be \SI{1}{\nano\meter} thick for simplicity of cross-material comparison.

~\sfref{fig:scheme}{e} shows the tunability of $pr^\text{MA}_\text{target}$ via the choice of different SU-8 height.
As the SU-8 height decreases from \SI{10}{\micro\meter} to \SI{2}{\micro\meter}, $pr^\text{MA}_\text{target}$ increases from $0.660 \times10^{-6}$ to $14.8\times10^{-6}$, while the $pr^\text{MA}_\text{flat}$ ($pr^\text{MA}_\text{side}$) increase from $2.52(1.99) \times10^{-6}$ to $9.36(2.78) \times10^{-6}$. 
Consequently, the ratio
$pr_{\mathrm{target}}^{\mathrm{MA}}/
 pr_{\mathrm{side}}^{\mathrm{MA}}$
increases from $0.332$ to $5.32$, corresponding to approximately $16$ times greater sensitivity
to surface-oxide loss in the target film.
For \SI{4}{\micro\meter} SU-8 height used in this experiment, $pr^\text{MA}_\text{target}=4.3\times10^{-6}$.

\section{Superconductor thin film loss}

\subsection{Baseline flip-chip-based sensor performance}\label{app:baseline}
To assess the loss of different target superconductor thin films, we use a \SI{60}{\nano\meter} sputtered Nb on Si film as a calibration film.
The film is grown under the same conditions as Ref.~\cite{olszewski_low_2025} and exhibits $\delta^\text{LP}\approx 1.5\times 10^{-6}$ performance on a 3-6-3 CPW resonator without acid-treatment.

\begin{figure}
    \centering
    \includegraphics[width=0.95\linewidth]{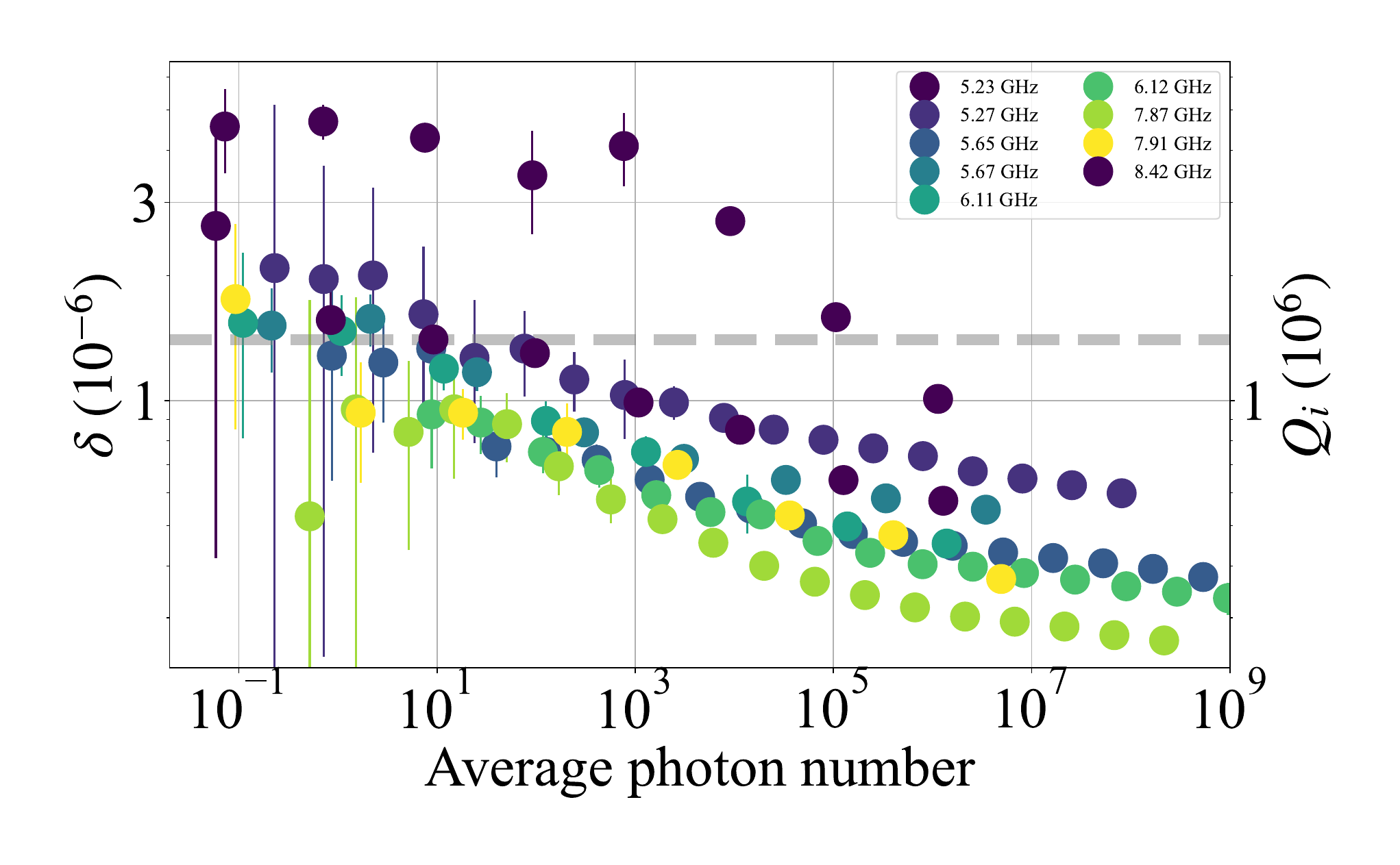}
    \caption{\textbf{Baseline flip-chip-based sensor performance.}
    Internal loss ($\delta$) average photon number dependence under \SI{60}{\nano\meter} Nb target film, deposited on Si substrate.
    The baseline $\delta^\text{LP}$ of the sensing resonator is $(1.5\pm1.0)\times 10^{-6}$, indicated by the dashed-gray line.
    The dataset consists of data from both sensors involved in the study and shows good consistency across them.
}
    \label{sfig:baseline}
\end{figure}

In \fref{sfig:baseline}, we show the power-dependence curves of the flip-chip-based sensor with a target Nb film.
Data from two sensors are shown together and exhibit good agreement.
From these data, we extract $\delta_\text{baseline}^\text{LP}=(1.5\pm1.0)\times10^{-6}$ and $\delta_\text{baseline}^\text{HP}= (0.48\pm0.28)\times10^{-6}$.

\subsection{Estimation of surface oxide loss}\label{app:conductor}

\begin{figure}
    \centering
    \includegraphics[width=0.95\linewidth]{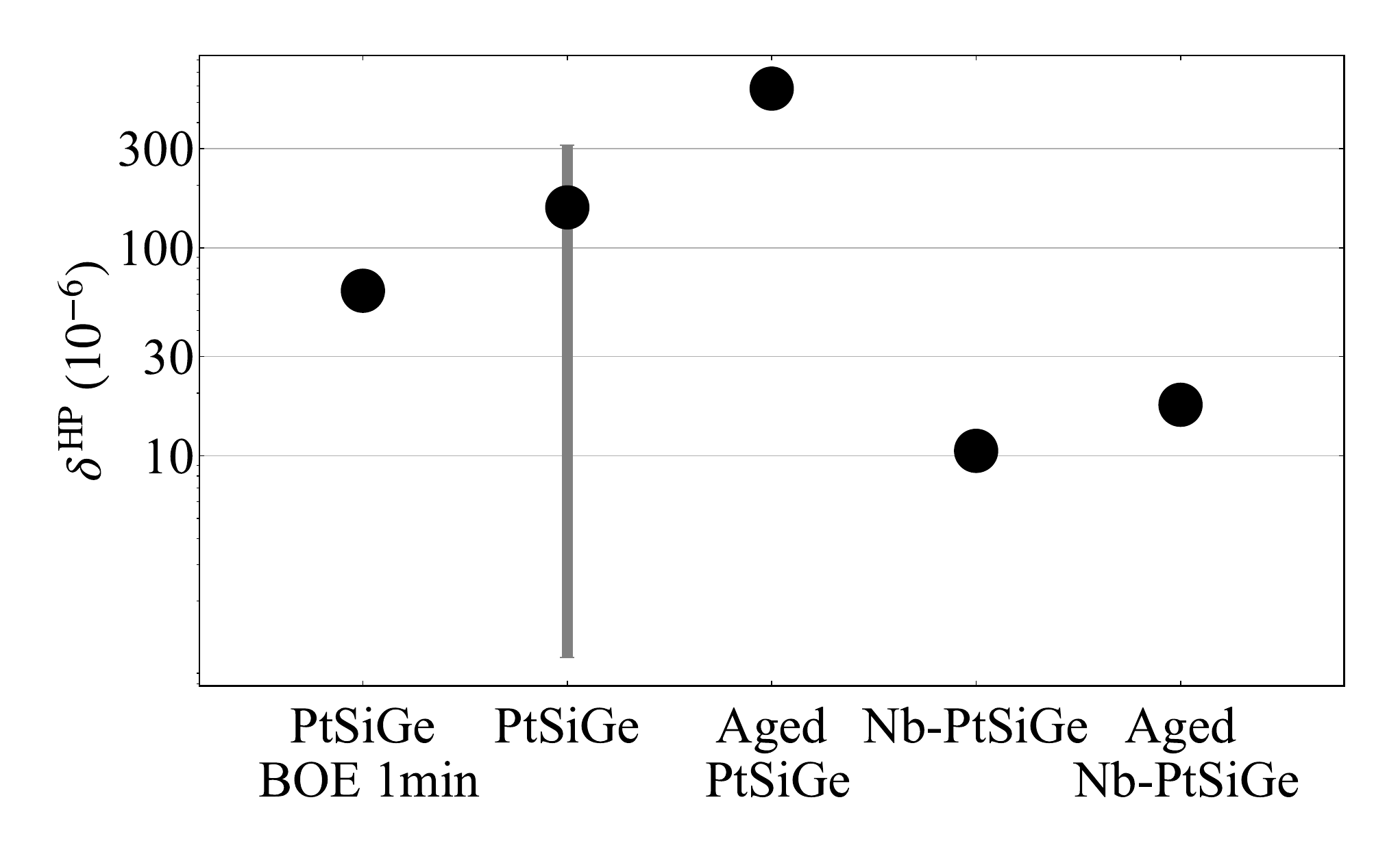}
    \caption{\textbf{$\delta^\text{HP}$ of different superconductor thin films subtracting the baseline.}
    The statistical uncertainty is defined as $\sqrt{(\sigma^2_\text{target})+(\sigma^2_\text{baseline})}$, where $\sigma$ is the standard deviation of $\delta^\text{HP}$ of the target material or baseline (Nb on Si).
    }
    \label{sfig:conductor}
\end{figure}

The high-power loss $\delta^\text{HP}$ of the different target films is shown in \fref{sfig:conductor}.
We observe a clear trend of $\delta^\text{HP}$ increasing over time for both PtSiGe and Nb-PtSiGe.
The aging of Nb-PtSiGe is more pronounced that of the Nb films reported in Ref.~\cite{olszewski_low_2025}, indicating that a Nb overlayer on PtSiGe does not afford the same aging resistance as Nb grown directly on Si.

For the three films discussed here we measure high-power losses $\delta^\text{HP} = (62.6\pm9.3)$, $(157\pm155)$ and $(11.0\pm1.4)\times10^{-6}$, and low-power losses $\delta^\text{LP} = (65.3\pm9.7)$, $(158\pm141)$ and $(13.2\pm2.0)\times10^{-6}$, for BOE-treated PtSiGe, as-grown PtSiGe and as-grown Nb-PtSiGe, respectively.

The STEM-EELS maps (\sfref{fig:TEM}{c,d}) show thick oxides at the metal-air interface of both PtSiGe and Nb-PtSiGe films, motivating the hypothesis that the measured low-power loss originates in surface dielectric at the interface.
If the entire excess low-power loss were attributed to the target metal-air interface, Eq.~\eqref{app:di_loss} would require a loss tangent
\begin{equation}
\tan\delta^\text{MA}_\text{req}
= \frac{\delta^\text{LP} - \delta^\text{LP}_\text{baseline}}{pr^\text{MA}_\text{target}},
\label{eq:tanreq}
\end{equation}
with $pr^{\rm MA}_{\rm target} = 4.3\times10^{-6}$ simulated for a $1\,$nm layer of $\epsilon_r = 10$ (\aref{app:MA}).

Evaluating this for the three films gives $\tan\delta^\text{MA}_\text{req} = 14.8\pm2.3$, $36.4\pm32.8$ and $2.72\pm0.52$, each of which substantially exceeds unity. 
In case of Nb-PtSiGe, the surface oxides are approximately \SI{5}{\nano\meter}, and the relative permittivity of Nb$_2$O$_5$ is around 40. 
Assuming linearity of the MA losses~\cite{gimbutas_precise_2026}, we can evaluate the $pr_\text{target}^{MA}$ for arbitrary thickness $t$ and dielectric constant $\epsilon_r$ as
\begin{equation}
pr^\text{MA}_\text{target}(t, \epsilon_r) =pr^\text{MA}_\text{target} \frac{t}{t_0} \frac{\epsilon_{r,0}}{\epsilon_r}
\label{eq:pr}
\end{equation}
where $t_0 = \SI{1}{\nano\meter}$ and $\epsilon_{r,0} = 10$. This brings the estimated MA losses to $\tan\delta^\text{MA}_\text{req} \sim 2.2$.
This is inconsistent by several orders of magnitude with our measurements for pure Nb flims and previous experiments on Nb\textsubscript{2}O\textsubscript{5}~\cite{Goronzy2025}.
Similarly, as-grown PtSiGe has a surface oxide layer which is approximately \SI{2}{\nano\meter} according to TEM (about two weeks older than the resonator sample), and predominantly composed of SiO$_x$ and GeO$_x$. Amorphous SiO$_x$ has a dielectric constant generally in the range of 3.9-4.5, and GeO$_x$ is typically 5.5-6.5. For an order-of-magnitude estimate, we assume the dielectric constant of this layer to be 5, which yields  $\tan\delta^\text{MA}_\text{req} \sim 42.7$ for the aged PtSiGe film.
These quantities therefore suggest that metal-air dielectric loss alone cannot account for the observed low-power loss.

We can therefore consider other interpretations. 
The near-equality of $\delta^\text{LP}$ and $\delta^\text{HP}$ for all three films, the apparent metal-air loss tangent exceeding unity, and the temperature dependence of $\delta^\text{HP}$ described by Mattis-Bardeen theory with a reduced gap (\sfref{fig:metal}{b}), together, indicate that the microwave loss of these films is dominated by a power-independent channel that is not dielectric in origin.
The most likely candidate is residual conduction associated with quasiparticles in low-gap or incompletely proximitized regions, consistent with the reduced $T_C^{\rm RF}$ extracted in Sec.~\ref{sec:sc_film_loss}.
A study targeting conductor loss is left to future work.

\section{Scanning transmission electron microscopy}
To avoid the formation of intermetallic Ga species with the films that may introduce artifacts, STEM samples were prepared using a TESCAN Amber X2 plasma FIB-SEM with Xe plasma. Following a typical focused-ion-beam lift-out procedure, inverted cross-sectional lamellae were prepared and polished at low voltage. All data were acquired at 120 kV with a probe convergence angle of 30 mrad on an aberration-corrected Thermo Fisher Scientific (TFS) Spectra 300 X-CFEG. EELS data were collected at a camera length of 29.5 mm on a Thermo Fisher Selectris-X detector. Surface oxide formation of a few nanometers on the lamella is common during transport to the microscope and may be seen in the EELS spectra of thin sample regions.

\section{X-ray Photoelectron Spectroscopy (XPS) scan of as-grown PtSiGe and BOE-treated PtSiGe}\label{app:xps}
The impact of a \SI{1}{\minute} BOE treatment on an as-grown PtSiGe film was characterized with XPS using a ThermoScientific Nexsa G2 Surface Analysis system with a chamber pressure of $2\times10^{-7}$ Torr. The tool uses a monochromated Al K-\textalpha\;X-ray source, with the Fermi level calibrated with a silver standard. The scans were performed with \SI{100}{\micro\meter} spot size with a \SI{0.4}{\electronvolt} energy resolution for the general survey scan and \SI{0.1}{\electronvolt} energy resolution for the core-level spectra. Fitting was done with a Shirley background in CasaXPS \cite{FAIRLEY2021} based on calibrated sensitivity factors provided by Thermo Fisher Scientific. Pt and Si are fit using asymmetric Voigt-like line shapes. Ge and all Pt, Si, and Ge oxides are fit using symmetric line shapes.

Two PtSiGe target films were diced out of the same chip and measured via the flip-chip-based sensor in a dilution fridge.
One was treated with a 1 min BOE dip prior to measurement. 
The two chips went through the same fridge cycle and XPS scans were promptly carried out after unloading.
The fridge warm-up period was roughly 24 hours under high vacuum.

\begin{table}
    \centering
    \begin{tabular}{|c|c|c|c|c|c|}\hline
         &  C&  O&  Pt&  Si& Ge\\\hline
         No BOE At. \%&  18.5\%&  43.0\%&  17.0\%&  16.5\%& 5.0\%\\ \hline
         BOE At. \%&  20.9\%&  25.9\%&  27.1\%&  8.7\%& 17.4\%\\\hline 
    \end{tabular}
    \caption{\textbf{Survey scan of PtSiGe with and without BOE treatment.} Relative atomic percentage of elements found in XPS survey scan comparing the BOE-treated sample and the BOE-untreated sample. Si composition is significantly lowered after BOE, implying the removal of Si with its oxides due to the process. Consequently, other element composition occupies a higher percentage composition.
    }
    \label{tab:XPS_survey}
\end{table}

Based on the survey scan result in \tref{tab:XPS_survey}, the near-surface region indicates that BOE reduces the proportion of Si and O. Consequently, Pt and Ge become a larger proportion of the composition. 
The C composition also rose slightly.

\begin{table}[htb]
    \centering
    \begin{tabular}{|c|c|c|c|c|}
        \hline
        & \multicolumn{2}{c}{No BOE}\vline & \multicolumn{2}{c}{BOE}\vline\\ \cline{2-5}
        & B.E. (eV)& At. \%& B.E. (eV)&At. \%\\ \hline\hline
        Pt& 72.47& 95.8\%& 72.31&93.9\%\\ \hline
        PtO& 74.47& 4.2\%& 74.36&6.1\%\\ \hline\hline
        Si& 100.32& 25.1\%& 100.21&45.3\%\\ \hline
        SiO\textsubscript{2}& 103.13& 74.9\%& 102.53&54.7\%\\ \hline\hline
        Ge& 1217.79& 85.7\%& 1218.02&51.9\%\\ \hline
        GeO& 1219.33& 12.0\%& 1219.26&42.0\%\\ \hline
        GeO\textsubscript{2}& 1222.44& 2.3\%& 1221.58&6.1\%\\ \hline
    \end{tabular}
\caption{\textbf{Core spectra scans of elements with and without BOE treatment.} The peaks measured are Pt 4f, Si 2p, and Ge 2p, and reported binding energies for doublet peaks refer to the lower of the doublet peaks. Si shows significant decrease in oxide composition while Ge has a higher oxide composition after BOE treatment. Fit shapes for Pt did not provide a strong fit and the small difference in oxide might be confounded by fitting deviations.}
    \label{tab:XPS_oxides}
\end{table}

The core-level spectral fits in \tref{tab:XPS_oxides} show in detail the effect of BOE treatment on the respective oxides. The Si oxide fraction upon BOE treatment decreases by 20.2\% (from 74.9\% to 54.7\%), corresponding to a 27\% relative decrease. On the other hand, Ge oxides increased in content by over 30\%, likely due to new exposure to atmosphere post-removal of Si oxides. PtO content did not change significantly, and due to difficulties in Pt peak fitting, we withhold any conclusions about the effect of BOE  on Pt in PtSiGe. 

\section{Other works}
A summary of measured microwave losses across different works is presented in \tref{table:comparison}.

\begin{table*}[htb]
    \caption{Compendium of microwave losses on Ge-based heterostructures. $^1$Resonator fabricated on Si substrate. $^2$QW removed by etching.}
    \label{table:comparison}
    \centering
    \begin{tabular}{c c c c c c c c}
        \hline\hline
        Reference & Substrate & Growth & Heterostructure & Metal & $f_r$ & LP loss & HP loss \\
        & & method & (QW/Spacer) & & (GHz) & ($10^{-6}$) & ($10^{-6}$) \\
        \hline
        \cite{Nigro_loss_QW_2024} & Si & CVD & \SI{15}{\nano\metre} Ge / & Al & 5--7 & 1000--1250 & 1000--1250 \\
         & & & \SI{55}{\nano\metre} Si$_{0.2}$Ge$_{0.8}$ & & & & \\
        \hline
        \cite{valentini_parity_2024} & Si & Low-energy & \SI{18}{\nano\metre} Ge / & Al & 4--6 & 132--608 & 28--340 \\
        & & PECVD & \SI{5}{\nano\metre} Si$_{0.3}$Ge$_{0.7}$ & & & & \\
        \hline
        \cite{valentini_parity_2024} & Si & Low-energy & \SI{18}{\nano\metre} Ge / & Al & 4--6 & 224--575 & 152--450 \\
        & & PECVD & \SI{60}{\nano\metre} Si$_{0.3}$Ge$_{0.7}$ & & & & \\
        \hline
        \cite{valentini_parity_2024} & Si & Low-energy & No Ge QW & Al & 4--6 & 100--370 & 33--111 \\
        & & PECVD & & & & & \\
        \hline
        \cite{Ruggiero2026} & Ge & MBE & \SI{15}{\nano\metre} Ge / & Al & 4--7 & 50--1000 & 20--1000 \\
        & & & \SI{5}{\nano\metre} Si$_{0.05}$Ge$_{0.95}$ & & & & \\
        \hline
        \cite{Palma_preprint_2025} & Si & CVD & \SI{16}{\nano\metre} Ge / & Al-AlOx & 5--7 & 250--333 & -- \\
        & & & \SI{60}{\nano\metre} Si$_{0.2}$Ge$_{0.8}$$^1$ & junction array & & & \\
        \hline
        \cite{Palma_preprint_2025} & Si & CVD & No Ge QW$^1$ & Al-AlOx & 5--7 & 50--100 & -- \\
        & & & & junction array & & & \\
        \hline
        \cite{Palma_preprint_2025} & Si & CVD & No Ge QW$^1$ & Nb & 5--6 & 10 & 2.5 \\
        \hline
        \cite{Jank2025} & Si & Low-energy & No Ge QW$^2$ & grAl & 7.26 & 500 & -- \\
        & & PECVD & & & & & \\
        \hline
        \cite{kiyooka_gatemon_2025} & Si & Reduced-pressure & No Ge QW$^2$ & Al & 6.185 & 68 & -- \\
        & & CVD & & & & & \\
        \hline
        \cite{Ruggiero_prerprint_2026} & Ge & MBE & \SI{15}{\nano\metre} Ge / & Al & 4.665 & 89 & -- \\
        & & & \SI{5}{\nano\metre} Si$_{0.05}$Ge$_{0.95}$ & & & & \\
        \hline\hline
    \end{tabular}
\end{table*}
\newpage

\section{Fridge wiring}

\begin{figure*}[htb]
\centering
\includegraphics[width = \textwidth]{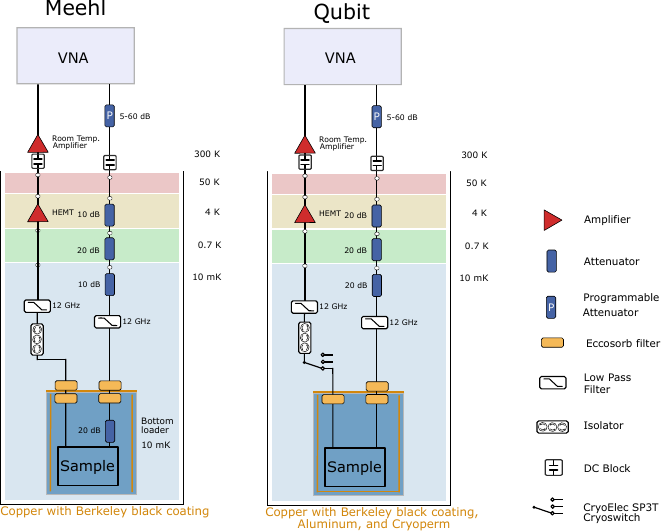}
\caption{ 
\textbf{Fridge wiring diagram.} 
}
\label{sfig:fridge_wiring} 
\end{figure*}

The samples were measured in two Bluefors dilution fridges with $\approx \SI{10}{\milli\kelvin}$ base temperature, labeled \textbf{Qubit} and \textbf{Meehl}, for which we found no resonator performance difference from earlier work \cite{olszewski_low_2025}. 

The \textbf{Qubit} fridge is a Bluefors LD fridge.
The input side contains $\SI{60}{\deci\bel}$ attenuation from attenuators located at different stages and $\approx \SI{7}{\deci\bel}$ from coax cables, one KL filter, and one flange-mount Eccosorb filter.
The output side contains one flange-mount Eccosorb filter, one low-loss RF switch and one three-terminal isolator at the MXC chamber. 
The output signal is amplified by one $\approx \SI{37}{\deci\bel}$ gain HEMT at 4K stage and one room-temperature $\approx \SI{38}{\deci\bel}$ gain amplifier. 

The \textbf{Meehl} fridge is a Bluefors LD fridge with a FSE bottom loader system.
The sample is placed in a gold-coated copper probe, enclosed by a Berkeley black-painted copper shim mounted between the probe and the sample.
The input side contains $\SI{60}{\deci\bel}$ attenuation from attenuators located at different stages and $\approx \SI{7}{\deci\bel}$ from coax cables.
The RF signal is connected via SMP connectors mounted on the MXC flange between the probe and the main body of the fridge.
Eccosorb filters are mounted on both sides of the SMP connector to suppress stray photons.
The output side contains one low-loss RF switch and one three-terminal isolator at the MXC chamber. 

We used a Copper Mountain M5180 VNA to collect the resonator data.
A RUDAT-13G-60 programmable attenuator of a $\SI{60}{\deci\bel}$ tuning range was connected to the signal output side of the VNA, enabling automatic script-based scan.

The schematic of the fridge wiring is depicted in ~\fref{sfig:fridge_wiring}.
\clearpage
\bibliographystyle{apsrev4-2}
\bibliography{references.bib}
\end{document}